\documentclass[english,nofootinbib]{revtex4-2}
\usepackage[T1]{fontenc}
\usepackage[latin9]{inputenc}
\usepackage{color}
\usepackage{array}
\usepackage{mathtools}
\usepackage{dsfont}
\usepackage{multirow}
\usepackage{amsmath}
\usepackage{stackrel}
\usepackage{graphicx}

\makeatletter

\providecommand{\tabularnewline}{\\}

\usepackage{babel}

\makeatother

\usepackage{babel}
\begin{document}
\title{Use of Nash equilibrium in finding game theoretic robust security
bound on quantum bit error rate }
\author{Arindam Dutta}
\email{arindamsalt@gmail.com}

\email{https://orcid.org/0000-0003-3909-7519}

\author{Anirban Pathak}
\email{anirban.pathak@gmail.com}

\email{https://orcid.org/0000-0003-4195-2588}

\affiliation{Department of Physics and Materials Science \& Engineering, Jaypee
Institute of Information Technology, A 10, Sector 62, Noida, UP-201309,
India}
\begin{abstract}
Nash equilibrium is employed to find a game theoretic robust security
bound on quantum bit error rate (QBER) for DL04 protocol which is
a scheme for quantum secure direct communication that has been experimentally
realized recently. The receiver, sender and eavesdropper (Eve) are
considered to be quantum players (players having the capability to
perform quantum operations). Specifically, Eve is considered to have
the capability of performing quantum attacks (e.g., W{\'o}jcik's
original attack, W{\'o}jcik's symmetrized attack and Pavi{\v{c}}i{\'c}
attack) and classical intercept and resend attack. Game theoretic
analysis of the security of DL04 protocol in the above scenario is
performed by considering several game scenarios. The analysis
revealed the absence of a Pareto optimal Nash equilibrium point within
these game scenarios. Consequently, mixed strategy Nash equilibrium points
are identified and employed to establish both upper and lower bounds
for QBER. Further, the vulnerability of the DL04 protocol to Pavi{\v{c}}i{\'c}
attack in the message mode is established. In addition, it is observed
that the quantum attacks performed by Eve are more powerful than the
classical attack, as the QBER value and the probability of detecting
Eve's presence are found to be lower in quantum attacks compared to
classical ones.
\end{abstract}
\maketitle

\section{Introduction}

Game theory examines and models how individuals behave in situations
involving strategic (rational) thinking and interactive decision-making.
It is crucial for decision-making processes and assessing opportunities,
both in business and everyday scenarios. Instances requiring strategic
thinking are prevalent in fields such as economics \cite{G92}, political
science \cite{O86}, biology \cite{C13,NS99} and military applications
\cite{D59,H97}. Participants in these scenarios have their own sets
of potential actions, referred to as strategies, and express preferences
for these actions through a payoff matrix. Game theory is concerned
with representing these activities and identifying optimal strategies.
Among the various concepts in game theory, the Nash equilibrium is
particularly significant. It characterizes the optimal decisions considering
the actions of other players. In a Nash equilibrium, no player stands
to benefit by altering their strategy alone \cite{N50,N51}.

Quantum mechanics stands out as one of the most influential theories
throughout history. Despite the controversies it has sparked since
its inception, its predictions have been consistently and precisely
confirmed through experiments \cite{AGR82}.\textcolor{red}{{} }Quantum
game theory enables the examination of interactive decision-making
scenarios involving players utilizing quantum technology. This technology
serves a dual purpose: functioning as a quantum communication protocol
and providing a more efficient method for randomizing players' strategies
compared to classical games \cite{L11}. Quantum game theory emanated
in 1999 through the contributions of David Meyer \cite{M99}, and Jens
Eisert, Martin Wilkens and Maciej Lewenstein \cite{EWL99}. Their
research explored games that incorporated quantum information, showcasing
scenarios where quantum players demonstrated advantages over their
classical counterparts. Subsequently, numerous examples of quantum
games, largely built upon the foundations established by Meyer, and
Eisert, Wilkens and Lewenstein, have been extensively examined. Further
details and references can be found in surveys such as \cite{GZK08}.
In the realm of experiments, researchers have successfully implemented
the quantum version of the Prisoners' Dilemma game using an NMR quantum
computer \cite{DLX+02}. Additionally, Vaidman \cite{V99} demonstrated
a simple game wherein players consistently emerge victorious if they
share a GHZ state beforehand, contrasting with classical players where
winning is always subject to probability. Quantum strategies have
been employed to introduce fairness elements into remote gambling
scenarios \cite{GVW99} and in formulating algorithms for quantum
auctions, which come with numerous security advantages \cite{P07}.
Flitney and Abbott \cite{FA03} have explored quantum adaptations
of Parrondo's games. Such analyses not only aid in designing secure
networks leading to the identification of novel quantum algorithms
but also add an entirely distinct dimension to characterizing a game
or a protocol \cite{DP+23,DP2023,DP+24}. Furthermore, eavesdropping \cite{E91,NH97}
and optimal cloning \cite{W98} can be conceptualized as games played
between participants. Thus, an interconnection between game theory and
quantum mechanics is already explored. There are primarily two approaches
that illustrate this interconnection. In the first approach, quantum
resources are used to play a traditional game that can also be played
without any quantum resources, but the use of quantum resources provide
some advantages, whereas in the second approach, a quantum mechanical
scenario is described using the concepts of the game theory. In what
follows, the first (second) approach mentioned above is referred to
as the \emph{quantized game}\emph{(gaming
the quantum}). Here, it will be apt to note that
a quantized game can be defined as a unitary function mapping the
Cartesian product of sets of quantum superpositions of players' pure
strategies to the entire Hilbert space encompassing the domain of
the game. This retains the characteristics of the original game under
specific conditions. On the other hand, the concept of \emph{gaming
the quantum} \cite{KP+13} refers to the application(s)
of the principles of game theory to quantum mechanics to derive game-theoretic
solutions. In this study, we do ``gaming the quantum'' by applying
non-cooperative game theory to a quantum communication protocol to
demonstrate how Nash equilibrium can serve as a viable solution concept.

By drawing motivation from the aforementioned facts, we can leverage
Nash equilibrium points in a game to establish secure bounds for various
quantum information parameters. This involves transforming any quantum
scheme into a scenario resembling a game and examining Nash equilibrium
points. Nash equilibrium points serve as stable conditions or provide
optimistic and rational probabilities for stakeholders to make decisions
within a mixed strategic game framework. Utilizing these probabilities
derived from Nash equilibrium points facilitates the establishment
of a stable gaming environment for all parties involved. This enables
the assessment of various cryptography parameters, including determining
the threshold value of quantum bit error rate (QBER). In a practical
context where decisions are autonomously made by each party, attaining
a stable point is less probable and could result in increased instability.
Consequently, identifying the minimum QBER value from the stable gaming
scenario sets a secure threshold boundary for the practical deployment
of the protocols for secure quantum communication. In our paper, we
specifically delve into the investigation of the secure threshold
bound for the QBER in the context of the DL04 protocol \cite{DL04}.
It is crucial to emphasize the context of direct secure quantum communication
protocols as the DL04 scheme belongs to this category. The direct
secure quantum communication protocols can be broadly categorized
into two classes \cite{LDW+07}. First, there are the deterministic
secure quantum communication (DSQC) protocols \cite{ZXF+06,HHT11,DP22,DP23},
where the receiver can decode the secret message sent by the sender
only after transmission of at least one bit of additional classical
information for each qubit. Second, there are the quantum secure direct
communication (QSDC) protocols \cite{BEK+02,LL02,BF02,DLL03,DL04,DBC+04,WDL+05,ZSL20,WLY+19},
which do not necessitate any exchange of classical information.  Beige
et al. introduced a QSDC scheme \cite{BEK+02}. In this proposal,
the message is accessible only following the transmission of additional
classical information for each qubit. Bostr{\"o}m and Felbinger presented
a ping-pong QSDC scheme \cite{BF02}, which is secure for key distribution
and quasi-secure for direct secret communication when employing a
perfect quantum channel. In 2004, Deng and Long \cite{DL04} put forth
a QSDC protocol (DL04 protocol) that does not rely on entangled states.
An unexpected finding is the ability to ensure secure information
transmission through the two-photon component, aligning with the outcomes
observed in two-way quantum key distribution (QKD) \cite{DL+04,LM_05,L19}.
This aligns with the specific scenario presented in the DL04 QSDC
protocol \cite{DL04}. Our primary focus in this study is on estimating
the secure threshold bound of QBER within the DL04 protocol. This
choice is motivated by its widespread acceptance for experimental
realization and its feasibility for secure implementation \cite{HYJ+16,ZZS+17,QSL+19,ZSN+20,PLW+20,PSL23,NZL+18}.

The remainder of the paper is structured as follows. In Section \ref{sec:II},
we begin with a lucid introduction to game theory restricted to the
context of the present paper as game theory serves as the basis for
our analysis. In this section, we also define the DL04 protocol as
a scenario resembling a game. Moving on to Section \ref{sec:III},
we present the mathematical formulation of our game and employ a graphical
method to examine the Nash equilibrium. Additionally, we scrutinize
the Nash equilibrium points to determine the secure threshold bound
of QBER for the DL04 protocol. Finally, Section \ref{sec:IV} provides
a summary and discussion of our findings, serving as the conclusion
of the paper. In addition, detailed mathematical proofs and details
of the analysis presented in various sections are presented in Appendices
A-E.

\section{Preliminaries of Game Theory\label{sec:II}}

Before we delve into the technical details of our work, it will be
apt to briefly introduce the basics of \emph{game theory}. Game theory
comprises a set of mathematical models designed to examine scenarios
involving both competition and collaboration, where an individual's
(player's) ability to make choices effectively relies on the choices
made by others. Each player has a set of possible strategies or actions
they can take. This strategy is a plan or decision that specifies
how a player will act in different situations within the game. The
primary objective is to identify optimal strategies for individuals
facing such situations and to identify equilibrium states. The outcomes
of a game are often represented in terms of payoffs, which measure
the utility or satisfaction that each player receives based on the
chosen strategies of all players. Payoffs can be represented in various
forms, such as numerical values, rankings or other measures. Game
theory can be represented in different forms; the\emph{ normal form}
(strategic form) is a matrix that shows the payoffs for each combination
of strategies chosen by the players. The \emph{extensive form }(dynamic
game) uses a tree-like diagram to represent sequential and simultaneous
decision-making. There are mainly two types of games, \emph{zero-sum
games} and \emph{non-zero-sum games}. In a zero-sum game, the total
payoff is constant, and gains for one player result in losses for
the other player(s). In non-zero-sum games, the total payoff can vary,
and the interests of the players may not be directly opposed. An important
fundamental concept in game theory is Nash equilibrium which corresponds
to a set of strategies in which no player has an incentive to unilaterally
change their strategy, given the strategies chosen by the other players.
It represents a stable solution where no player can improve their
payoff by changing their strategy. A game's strategy set is considered
\emph{Pareto efficient} (or \emph{Pareto optimal}) when there does
not exist another strategy set that can improve the outcome for one
player without negatively affecting any other player. Further, a dominant
strategy is a strategy that is always the best choice for a player,
regardless of the strategies chosen by other players. A dominant strategy
is a strong concept in rational decision-making. When a player follows
a pure strategy, they choose a single action or decision without any
randomness or uncertainty. In some games, players may adopt mixed
strategies, where they choose their actions with certain probabilities.
Mixed strategies can lead to a Nash equilibrium when no pure strategy
is optimal \cite{EWL99,PS03,alonso2019quantum,BW22,KK18}.

\emph{Quantum game }In a traditional game that allows for the use
of mixed strategies, players construct their strategies by using real
coefficients to form convex linear combinations of their pure strategies
\cite{O04}. In contrast, in a quantum game \cite{EWL99,EW2000},
players employ unitary transformations and quantum states that belong
to substantially larger strategy spaces. This has led to discussions
proposing that quantum games could be considered extensions of classical
games, potentially offering stakeholders a quantum advantage in the
game \cite{EP02}. As mentioned earlier, it is important to emphasize
that any quantum protocol (quantum cryptographic scenarios) can be
analogized to a game-like scenario. In this context, participants'
strategies hinge on the utilization of unitary operations and the
selection of measurement bases, with the final payoff determined by
the results of the measurements. Quantum advantage arises from the
optimal sequential utilization of quantum operations on quantum states
by the players. Typically, in traditional quantum game theory paper
classical players are limited to using coherent permutations of standard
basis states, or similarly restricted types of unitary operations.
In contrast, quantum players have access to a wider range of available
unitary operations, which may include the full spectrum of such operations
with fewer limitations. Before
we proceed further, it will be appropriate to formally discuss \emph{pure
quantum strategy}, \emph{mixed
quantum strategy} and \emph{positive
operator valued measure (POVM) quantum strategy}
\cite{EW2000} in the context of the present work. A pure quantum
strategy involves deterministic actions, represented by unitary operations
on the player's quantum state. This strategy starts with a quantum
state $|\Psi\rangle$ in a Hilbert space $\mathcal{H}$. Each player
$i$ selects a unitary operation $U_{i}$ to apply to their portion
of the quantum state. After all players have applied their respective
unitary operations, the resulting quantum state is $|\Psi_{f}\rangle=\left(U_{1}\otimes U_{2}\otimes\cdots\otimes U_{i}\otimes\cdots\otimes U_{n}\right)$,
where $n$ denotes the total number of players in a game. The final
state $|\Psi_{f}\rangle$ is then measured to determine the outcome
of the game. On the other hand, a mixed quantum strategy involves
probabilistic combinations of various pure quantum strategies. Starting
with the same initial state as in the pure strategy, each player $i$
has a set of pure strategies $\left\{ U_{i}^{1},U_{i}^{2},\cdots,U_{i}^{k}\right\} $
and an associated probability distribution $\left\{ p_{i}^{1},p_{i}^{2},\cdots,p_{i}^{k}\right\} $,
where $\stackrel[j=1]{k}{\sum}p_{i}^{j}=1$. Each player $i$ randomly
selects a unitary operation $U_{i}$ with probability $p_{i}^{j}$.
Consequently, the ensemble of possible final states is a mixture of
pure states, represented by the mixed state $\rho_{f}=\underset{j}{\sum}p_{i}^{j}\left(U_{i}^{j}\otimes\cdots\otimes U_{n}^{j}\right)|\Psi\rangle\langle\Psi|\left(U_{i}^{j}\otimes\cdots\otimes U_{n}^{j}\right)^{\dagger}$.
This mixed state $\rho_{f}$ is then measured to determine the game's
outcome. A POVM quantum strategy employs general quantum measurements,
defined by a set of POVM elements. Starting with the initial state
$|\Psi\rangle$, each player $i$ uses a POVM, which consists of positive
semi-definite operators $\left\{ \mathcal{E}_{i}^{1},\mathcal{E}_{i}^{2},\cdots,\mathcal{E}_{i}^{k}\right\} $
that satisfy $\underset{j}{\sum}\mathcal{E}_{i}^{j}=\mathds{1}$.
The POVM elements are applied to the quantum state, with outcome $j$
occurring with probability $p_{i}^{j}=\langle\Psi|E_{i}^{j}|\Psi\rangle$.
Depending on the measurement outcome, players may apply conditional
unitary operations or other quantum operations. The final state, which
depends on the measurement outcomes and subsequent operations, is
given by $\rho_{f}=\underset{j}{\sum}\left(U_{i}^{j}\otimes\cdots\otimes U_{i}^{j}\right)\mathcal{E}_{i}^{j}|\Psi\rangle\langle\Psi|\mathcal{E}_{i}^{j}\left(U_{i}^{j}\otimes\cdots\otimes U_{i}^{j}\right)^{\dagger}$.
This final state $\rho_{f}$ is measured to determine the game's outcome.
Each type of strategy improves in generality and flexibility, with
POVM quantum strategies covering the widest range of possible actions
within quantum game theory. In our work, we have implemented a mixed
quantum strategy in which Alice, Bob and Eve perform their quantum
operations with specific probabilistic combinations to get Nash equilibrium
points. The final mixed state determines the payoff elements for each
respective player.

Our considered game, which is a 
non-cooperative quantum game involving multiple
players, lacks a pure strategy Nash equilibrium. This is due to the
nature of the multiplayer model and the way payoffs are calculated,
similar to the approach in \cite{KP+13}. Furthermore, the concept
of a Nash equilibrium solution is fundamental in game theory as it
can be used to predict the behavior of non-cooperating players. The
proof of Nash\textquoteright s theorem for the existence of an equilibrium
in mixed strategies in traditional games is relatively simple and
relies entirely on Kakutani\textquoteright s fixed-point theorem \cite{Kakutani1941}.
In the realm of quantum games, Meyer demonstrated the existence of
Nash equilibrium in mixed strategies, which are represented as mixed
quantum states, by employing Glicksberg\textquoteright s \cite{Glicksberg1952}
extension of Kakutani\textquoteright s fixed-point theorem to topological
vector spaces. In 2019, Khan et al. found that the Kakutani fixed-point
theorem does not directly apply to quantum games involving pure quantum
strategies \cite{KH19}. However, by using Nash\textquoteright s embedding
theorem, which embeds compact Riemannian manifolds into Euclidean
space \cite{Nash_1956}, and under certain conditions, the Kakutani
fixed-point theorem can be indirectly applied to ensure Nash equilibrium
in pure quantum strategies. They did a formal mathematical discussion
of non-cooperative game theory and fixed points (see Ref. \cite{KH19}
for details).

\emph{Our contribution} 
In our game, several key points should be noted as it is a \emph{mixed
quantum strategic game}. When players can choose
their quantum strategies based on a probability distribution, thereby
using mixed quantum strategies, Meyer demonstrated through Glicksberg\textquoteright s
fixed-point theorem \cite{Glicksberg1952} that a Nash equilibrium
will always be present. Meyer's research also offers insights into
the equilibrium behavior of quantum computational mechanisms. His
approach to identifying quantum advantage as a Nash equilibrium in
quantum games remains largely unexplored \cite{KSB+18}. Additionally,
in the realm of quantum communication protocols, where quantum processes
are typically noisy and represented as density matrices or mixed quantum
states, the Meyer--Glicksberg theorem ensures the existence of a
Nash equilibrium \cite{M99}. Generally, in quantum communication protocols, less restricted
quantum players can utilize their mixed strategies to calculate their
individual best response functions based on their payoffs. The solutions
of best response functions give the Nash equilibrium points. While
not all of these points may be Pareto optimal Nash equilibrium points,
they serve as a basis for determining the probabilities of mixed strategies
employed by the players. This, in turn, allows for an investigation
into the existence of Pareto optimal Nash equilibrium points. The
outcome of this analysis helps to determine the secure bounds for
various quantum information parameters (e.g., secret key rate, QBER)
with appropriate use of the fundamental concepts of quantum information
theory \cite{NC10,W17,W18,P13}. In our letter, we take the DL04 protocol
\cite{DL04} and analyze it as a game, considering different types
of attack (collective attacks and IR attack) to get the secure threshold
limit for QBER.

In quantum cryptography, QSDC employs quantum states as carriers of
information for secure communication. Unlike traditional methods,
QSDC does not require a prior generation of secret key \cite{BEK+02,LL02,BF02,DLL03,DL04,DBC+04,WDL+05}.
It is a concept centered around achieving secure and reliable communication
through the principles of quantum physics. Extensive experimental
studies on QSDC have demonstrated its feasibility and promising application
prospects \cite{HYJ+16,LHA+16,ZHS+17,QSL+19,MMB+19,PLW+20,ZSQ+22,LPS+22,PSL23}.
After more than two decades of persistent effort, QSDC is gradually
maturing and showing significant potential for advancing next-generation
secure communication \cite{YWH+21}, including potential applications
in military contexts \cite{K21}. Among different protocols for QSDC
that have been proposed till date, the DL04 protocol needs special
mention as the recent experimental activities \cite{HYJ+16,ZZS+17,QSL+19,ZSN+20,PLW+20,PSL23,NZL+18}
are centered around it. Keeping this in mind, in what follows we focus
our work on the DL04 protocol specifically though the strategy developed
here is general in nature.

Let us commence with a concise overview of the DL04 protocol \cite{DL04},
slightly modified to align with a gaming scenario. Within this framework,
Bob, at random, generates quantum states (photons) $|0\rangle$ and
$|1\rangle$ in the computational basis ($Z$ basis), and $|+\rangle$
and $|-\rangle$ in the diagonal basis ($X$ basis), with probabilities
$p$ and $1-p$, respectively. Subsequently, Bob transmits this sequence
to Alice. Alice operates in two modes: message mode (encoding mode)
and control mode (security check mode). In the control mode, Alice
randomly selects a subset of photons received from Bob to conduct
eavesdropping detection using a beam splitter. To complete the eavesdropping
detection, each photon in the subset is measured on either the $Z$
basis or $X$ basis. Alice communicates to Bob the positions of the
photons earmarked for security checks, the chosen measurement bases,
and the corresponding measurement results. Alice and Bob jointly estimate
the first QBER. If the QBER falls below a predefined threshold, they
proceed to the subsequent step; otherwise, the transmission is discarded.
In the message mode, Alice performs a quantum operation $I$ ($iY\equiv ZX$)
on the qubit state for the remaining photons to encode $0$ with probability
$q$ ($1$ with probability $1-q$). Additionally, she selects some
photons to encode random numbers $0$ or $1$ to assess the reliability
of the second transmission, which is then relayed back to Bob. The
majority of single photons in the second transmitting sequence carry
secret information, i.e., a message encoded by Alice, while a small
subset encodes random numbers for estimating the second QBER. Upon
receiving the photons from Alice, Bob decodes the classical information
bits encoded by Alice during the decoding phase, based on his preparation
bases. Alice discloses the positions of the photons encoding random
numbers, and both parties estimate the QBER of the second transmission.
The second QBER estimation primarily assesses the integrity of information
transmission, and if it falls below a specified threshold, the transmission
is deemed successful. A similar scheme, well-known as LM05 proposed
by Lucamarini et al. \cite{LM_05}, follows a similar process. In
this case, the control mode is executed by Alice akin to DL04, but
classical announcement is permitted at that stage of the protocol.
Upon receiving the sequence from Alice, Bob performs the decoding
phase through projective measurement and subsequently announces the
classical information of the security check bits. In both schemes,
legitimate users aspire to achieve a flawless \textquotedbl double
correlation\textquotedbl{} of measurement results on both the forward
and backward paths.

In the DL04 protocol, two quantum players employ strategies to ensure
secure communication. Our focus lies in determining the secure bound
on QBER when accounting for the presence of an eavesdropper (Eve),
who may utilize both quantum and classical attack strategies. We consider
that the quantum Eve employs the following attack strategies: W{\'o}jcik's
original attack \cite{W03}, W{\'o}jcik's symmetrized attack \cite{W03}
and Pavi{\v{c}}i{\'c} attack \cite{P+13}, denoted as $E_{1}$, $E_{2}$
and $E_{3}$, respectively. Additionally, there is a classical Eve
employing an Intercept Resend (IR) attack strategy ($E_{4}$). The
detailed analysis of attack strategies is conducted in \textcolor{black}{Appendices
A, B, C and D}. 
Here, we briefly describe Eve's attack strategies to aid readers in
understanding the concepts (for more details, see the Appendices).
In the $E_{1}$ attack, Eve utilizes the unitary operation $Q_{txy}$
during the B-A attack and its conjugate, $Q_{txy}^{\dagger}\left(\equiv Q_{txy}^{-1}\right)$,
during the A-B attack, where $Q_{txy}(={\rm SWAP}_{tx}\,{\rm CPBS}_{txy}\,H_{y}\equiv{\rm SWAP}_{tx}\otimes I_{y}\,{\rm CPBS}_{txy}\,I_{t}\otimes I_{x}\otimes H_{y})$.
Here, A-B and B-A attack refer to Eve's attack on the quantum channel
when state transmission is from Alice to Bob and from Bob to Alice,
respectively. This attack results in the state being in a higher dimensional
Hilbert state with Eve's unitary operations, leading to a higher degree
of randomization through quantum superposition. This affects the final
joint probabilities of Alice's, Bob's and Eve's measurement outcomes
as $p_{000}^{E_{1}}=q,$ $p_{001}^{E_{1}}=p_{010}^{E_{1}}=p_{011}^{E_{1}}=0,$
$p_{100}^{E_{1}}=\frac{1}{4}\left(1-q\right),$ $p_{101}^{E_{1}}=\left(1-q\right)\left(\frac{1}{4}+\frac{p}{2}\right),$
$p_{110}^{E_{1}}=0$ and $p_{111}^{E_{1}}=\frac{1}{2}\left(1-p\right)\left(1-q\right)$.
The QBER in the $E_{1}$ attack is $\frac{1}{2}\left(1-q\right)\left(1+p\right)$
and the detection probability of Eve's presence is 0.1875. The $E_{2}$
attack is similar to the $E_{1}$ attack, with the only difference
being that with a probability of $\frac{1}{2}$, an additional unitary
operation $S_{ty}$ is applied right after the operation $Q_{txy}^{-1}$
during the A-B attack. The $S_{ty}$ operation is defined as $X_{t}Z_{t}{\rm CNOT}_{ty}X_{t}$.
In this scenario, the final joint probabilities of Alice's, Bob's
and Eve's measurement outcomes are $p_{000}^{E_{2}}=\frac{1}{2}\left[\frac{q}{4}\left(1+p\right)+q\right]=\frac{q}{8}\left(5+p\right),$
$p_{001}^{E_{2}}=\frac{q}{8}\left(1+p\right),$ $p_{010}^{E_{2}}=p_{011}^{E_{2}}=\frac{q}{8}\left(1-p\right),$
$p_{100}^{E_{2}}=\frac{1}{4}\left(1-q\right),$ $p_{101}^{E_{2}}=\frac{1}{4}\left(1-q\right)\left(1+2p\right),$
$p_{110}^{E_{2}}=0$ and $p_{111}^{E_{2}}=\frac{1}{2}\left(1-p\right)\left(1-q\right)$.
The QBER in the $E_{2}$ attack is $\frac{1}{4}\left(2+2p-q-3pq\right)$
and the detection probability of Eve's presence is 0.1875. In the
$E_{3}$ attack, similar to the previous attacks, Eve applies the
unitary operation $Q_{txy}^{\prime}$ when the photon is traveling
from Bob to Alice in the B-A attack scenario. Conversely, in the A-B
attack scenario, Eve applies the inverse of $Q_{txy}^{\prime}$ $\left(Q_{txy}^{\prime-1}\right)$
to the traveling photon ($t$), where $Q_{txy}^{\prime}={\rm CNOT}_{ty}\left({\rm CNOT}_{tx}\otimes I_{y}\right)\left(I_{t}\otimes{\rm PBS}_{xy}\right){\rm CNOT}_{ty}\left({\rm CNOT}_{tx}\otimes I_{y}\right)\left(I_{t}\otimes H_{x}\otimes H_{y}\right)$.
In this scenario, the final joint probabilities of Alice's, Bob's
and Eve's measurement outcomes are $p_{000}^{E_{3}}=q,$ $p_{001}^{E_{3}}=p_{010}^{E_{3}}=p_{011}^{E_{3}}=0,$
$p_{100}^{E_{3}}=p_{101}^{E_{3}}=p_{110}^{E_{3}}=0,$ and $p_{111}^{E_{3}}=\left(1-q\right)$.
The QBER in the $E_{3}$ attack is 0 and the detection probability
of Eve's presence is 0.1875. The $E_{4}$ attack is essentially an
IR attack, where the final joint probabilities of Alice's, Bob's,
and Eve's measurement outcomes are $p_{000}^{E_{4}}=\frac{3}{4}q,$
$p_{001}^{E_{4}}=p_{011}^{E_{4}}=0,$ $p_{010}^{E_{4}}=\frac{1}{4}q,$
$p_{100}^{E_{4}}=p_{110}^{E_{4}}=0,$ $p_{101}^{E_{4}}=\frac{1}{4}\left(1-q\right)$
and $p_{111}^{E_{4}}=\frac{3}{4}\left(1-q\right)$. The QBER in the
$E_{4}$ attack is 0.25 and the detection probability of Eve's presence
is 0.375.

\begin{figure}[h]
\begin{centering}
\includegraphics[scale=0.55]{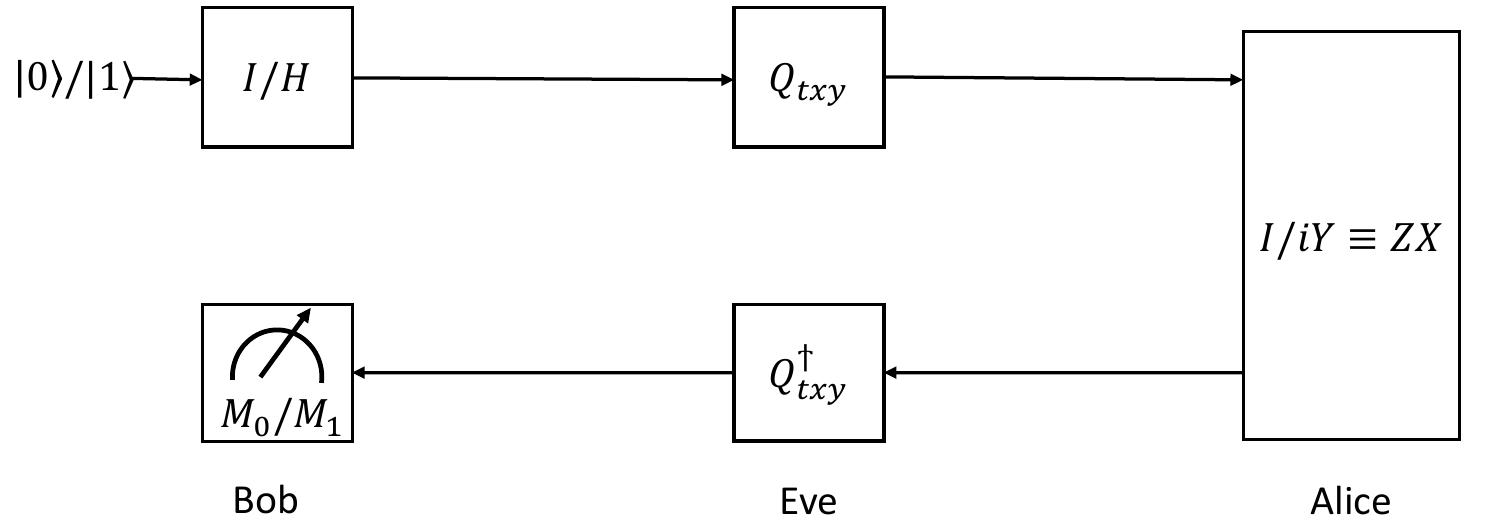}
\par\end{centering}
\caption{\label{fig:Circuit_Representation_DL04_QuantumGame}The
circuit representation of DL04 quantum game.}

\end{figure}

Before describing
the conventional payoff function, it is crucial to map the DL04 protocol
in a game-theoretic scenario where principles of game theory is applied
into a quantum communication protocol (see Figure \ref{fig:Circuit_Representation_DL04_QuantumGame}).
For simplicity, we divide DL04 quantum game into different game scenarios.
However, in what follows, we explain the mapping in a general context
\cite{EW2000,EWL99,KP13}. Let us consider game $\mathcal{G}$ as
a\emph{ normal form game}. This
game is defined as a function with an appropriate domain and range,
with players' preferences defined over the elements of the range,
inducing rational strategic choices within the domain. We now describe
the players' choices in the quantum domain. In more general terms,
the game $\mathcal{G}$ has a range that is equal to the output set
$O$ and a domain that is the Cartesian product $S_{1}\times S_{2}\times\cdots\times S_{n}$,
where $S_{i}$ represents the set of pure strategies for the $i^{th}$
player. Each element of the domain set $S_{1}\times S_{2}\times\cdots\times S_{n}$
can be referred to as a\emph{ play of game}
or a \emph{strategy profile}.
Formally, a game $\mathcal{G}$ is a function, $\mathcal{G}:S_{1}\times S_{2}\times\cdots\times S_{n}\longrightarrow O$.
In this context, $n$ represents the number of players. Since our
game involves three participants, Alice, Bob and Eve. Bob's choice
of prepared state, Alice's choice of measurement basis and Eve's choice
of attack strategy are within the domain set $S_{1}\times S_{2}\times S_{3}$,
where $S_{1},S_{2}$ and $S_{3}$ are the strategy choices of Alice,
Bob and Eve, respectively. Thus, the function $\mathcal{G}$ can be
specifically expressed as $\mathcal{G}:S_{1}\times S_{2}\times S_{3}\longrightarrow O$
in the form of pure strategies. Here, Bob prepares quantum states
in the $Z$ or $X$ basis and sends them to Alice. For a security
check, Alice measures part of Bob's sequence and then encodes a message
by performing the Pauli operation before sending it back to Bob. During
this process, Eve employs various attack strategies through the quantum
channel. In what follows, we elucidate how this game is incorporated
into the quantum communication protocol.

Alice arranges her initial sequence as a mixed state,
represented by $\frac{p}{2}\left(|0\rangle\langle0|+|1\rangle\langle1|\right)+\frac{1-p}{2}\left(|+\rangle\langle+|+|-\rangle\langle-|\right)$.
The states of Alice resides within a 2-dimensional projective complex
Hilbert space, denoted as $\mathcal{H}_{2}$. This mixed state contains
with four pure states, which can be interpreted as a superposition
state within $\mathcal{H}_{2}$. This implies that if $|0\rangle$
undergoes a projective measurement in the $X$ basis\footnote{Let us assume that the process of quantum measurement
in a 2-dimensional space can be characterized by the operator set
\{$M_{0}$,$M_{1}$: $M_{0}=|+\rangle\langle+|$, $M_{1}=|-\rangle\langle-|$\}.}, it will collapse into $|+\rangle$ and $|-\rangle$
with a probability of $\frac{1}{2}$. This occurrence is a result
of quantum superposition. Eve also executes attacks using quantum
unitary gates and ancilla states, which operate within the projective
complex Hilbert space. Specifically, the Eve's $E_{1}$, $E_{2}$
and $E_{3}$ attacks operate within an 8-dimensional Hilbert space
$\mathcal{H}_{8}$, and the $E_{4}$ attack operates within $\mathcal{H}_{2}$
space. In our different game scenarios, we assume that Eve attacks
with $E_{i}$ and $E_{j}$ with probabilities $r$ and $1-r$, respectively.
As a result of these operations by Eve, the states of Alice and Bob
are further increase the quantum
superposition based on the unitary operators utilized in Eve's attacks
( inplying that the basis used by alice total number of possible outcome
increases). This increase of superposition impacts the measurement
outcome of the legitimate player in this game scenario\footnote{The detailed descriptions of Alice's and Bob's unitary
operations and the results of their measurements are elaborately presented
in the appendices.} that also increase the range of the game $\left(O\right)$,
for $O=\left\{ o_{1},o_{2},o_{3}\right\} $, $o_{1},o_{2}$ and $o_{3}$
are the output of Alice, Bob and Eve, respectively. Alice employs
Pauli operations to encode the message, which also exists within $\mathcal{H}_{2}$
space. Consequently, we can describe our game as a mixed strategy
game, defined as a function $\mathcal{M}$. The 1$-$simplex $\Delta_{1}$,
represents the set of probability distributions over each player's
pure strategies, also known as the \emph{set of
mixed strategies}. Thus, our mixed strategy game
$\mathcal{M}$ is defined with a domain equal to the Cartesian product
of the sets of probability distributions over the pure strategies
of the players. Formally, this is represented as:

\[
\begin{array}{lcl}
\mathcal{M} & : & \Delta_{1}\times\Delta_{1}\times\Delta_{1}\longrightarrow\Delta_{7}\\
\\
\mathcal{M} & : & \left(\left(p,1-p\right),\left(q,1-q\right),\left(r,1-r\right)\right)\\
 & \mapsto & \left(pqr,pq\left(1-r\right),p\left(1-q\right)r,p\left(1-q\right)\left(1-r\right),\left(1-p\right)qr,\right.\\
 &  & \left.\left(1-p\right)q\left(1-r\right),\left(1-p\right)\left(1-q\right)r,\left(1-p\right)\left(1-q\right)\left(1-r\right)\right)
\end{array},
\]
Its range includes probability distributions over the outcomes by
associating these outcomes with the vertices of the 7$-$simplex $\Delta_{7}$.
The required projective complex Hilbert space for $\mathcal{M}$ is
as follows:

\[
\begin{array}{lcl}
{\rm Bob's\,prepared\,state} & \longrightarrow & \mathcal{H}_{2}\\
{\rm Alice's\,measuremet\,basis} & \longrightarrow & \mathcal{H}_{2}\\
{\rm Alice's\,unitary\,operation} & \longrightarrow & \mathcal{H}_{2}\\
{\rm Eve's\,}E_{1},E_{2},E_{3}\,{\rm attack} & \longrightarrow & \mathcal{H}_{8}\\
{\rm Eve's\,}E_{4}\,{\rm attack} & \longrightarrow & \mathcal{H}_{2}
\end{array}.
\]
Additionally, we observe that the superposition states belong to Bob,
along with the unitary operations executed by Alice and Eve, thus
the state belongs within the higher dimensional complex Hilbert space
($\mathcal{H}_{8}$). This process incorporates a higher degree of
randomization via quantum superposition. The impact of such high-level
randomization reflects on the outputs of the legitimate players or
increase the range of this game. Accordingly, these output results
influence on the key factors of the players' payoff function, such
as mutual information, QBER and the detection of Eve's presence. Ultimately,
these effects on the payoff functions determine the Nash equilibrium
points for the different game scenarios. By employing quantum superposition
states and quantum unitary operations, we apply non-cooperative game
theory to the DL04 protocol and demonstrate that Nash equilibrium
serves as a viable solution concept. Hence, we integrate \emph{gaming
the quantum} in our work \cite{KP13,KSB+18}.

We will now define the payoff for each party involved.
The mutual information between Alice and Bob has a positive impact
on the payoffs of legitimate players (Alice and Bob), but it negatively
affects Eavesdropper's (Eve) payoff. Conversely, the mutual information
shared between Alice and Eve, as well as Bob and Eve, negatively affects
the payoffs of legitimate players. Additionally, legitimate players
benefit from the ability to detect Eve's presence\footnote{The contribution of both parameters is equally significant, contingent
upon the specific quantitative values of QBER and the detection probability
of Eve's presence ($P_{d}$) observed in both message and control
modes of the DL04 protocol.}, which improves their payoffs in this competitive scenario. On the
other hand, Eve's payoff increases with the acquisition of more information
from Alice and Bob but decreases as the mutual information between
Alice and Bob increases. Furthermore, Eve incurs penalties if she
is detected. Consequently, Eve's payoff is enhanced when the probability
of remaining undetected is higher. Additionally, Eve may employ various
quantum gates to collect information from Alice and Bob, and the greater
the number of gates she uses, the more it impacts her overhead, which,
in turn, adversely affects her payoff. As a result, the payoff can
be customized to reflect the various scenarios and benefits of the
players that we intend to analyze. The DL04 protocol, resembling a
game, can be conceptualized as a zero-sum game. We may formulate the
payoffs of Alice, Bob and Eve \cite{KK18} for a general attack strategy $\mathcal{E}$
as follows\footnote{We assume that the difficulty of preparing quantum states and performing
measurement operations is the same for Alice, Bob and Eve. Therefore,
we do not factor in this contribution when estimating the payoffs
for these parties.},

\begin{equation}
\begin{array}{lcl}
P_{A}^{\mathcal{E}}(p,q) & = & \omega_{a}I\left({\rm A,B}\right)-\omega_{b}I\left({\rm A,E}\right)-\omega_{c}I\left({\rm B,E}\right)+\omega_{d}\left(\frac{P_{d}+{\rm QBER}}{2}\right)\\
\\
P_{B}^{\mathcal{E}}(p,q) & = & \omega_{a}I\left({\rm A,B}\right)-\omega_{c}I\left({\rm A,E}\right)-\omega_{b}I\left({\rm B,E}\right)+\omega_{d}\left(\frac{P_{d}+{\rm QBER}}{2}\right)\\
\\
P_{E}^{\mathcal{E}}(p,q) & = & -\omega_{e}I\left({\rm A,B}\right)+\omega_{f}I\left({\rm A,E}\right)+\omega_{g}I\left({\rm B,E}\right)+\omega_{h}\left(1-\frac{P_{d}+{\rm QBER}}{2}\right)-\omega_{i}n_{1}-\omega_{j}n_{2}-\omega_{k}n_{3}
\end{array},\label{eq:Initial_Payoffs}
\end{equation}
where $\omega_{a},\,\omega_{b},\,\omega_{c},\,\omega_{d},\,\omega_{e},\,\omega_{f},\,\omega_{g},\,\omega_{h},\,\omega_{i},\,\omega_{j}$
and $\omega_{k}$ are all positive real numbers, and they are interpreted
as the factors or weights assigned to each component within the payoff
such that $\stackrel[m=a]{d}{\sum}\omega_{m}=1$ and $\stackrel[n=e]{k}{\sum}\omega_{n}=1$
where $m\in\{a,b,c,d\}$ and $n\in\{e,f,\ldots,k\}$ . Further, $I\left({\rm A,B}\right)$,
$I\left({\rm A,E}\right)$ and $I\left({\rm B},{\rm E}\right)$ are
the mutual information between Alice and Bob, Alice and Eve and Bob
and Eve, respectively. Here, $P_{d}$ is the probability of detection
of Eve's presence, $n_{1}$, $n_{2}$ and $n_{3}$ are the number
of single-qubit, two-qubit and three-qubit gates, respectively. Here,
it may be noted that in the payoff function described above, $P_{d}$
and $QBER$ are considered on equal footing because both $P_{d}$
and $QBER$ reveal the presence of Eve with a similar effect.

Assuming Eve
has unlimited quantum resources and perfect quantum gates, we can
set $\omega_{i}=\omega_{j}=\omega_{k}=0$. This implies she can use
as many quantum gates as needed for her attack without negatively
impacting her payoff function. To simplify the analysis, in what follows
we have considered the effect of each component in the payoff is the
same. Thus, in our consideration, $\omega_{a}=\omega_{b}=\omega_{c}=\omega_{d}=\omega_{e}=\omega_{f}=\omega_{g}=\omega_{h}=0.25$.
The Eq. (\ref{eq:Initial_Payoffs}) is modified as,

\begin{equation}
\begin{array}{lcl}
P_{A}^{\mathcal{E}}(p,q) & = & 0.25\times\left[I\left({\rm A,B}\right)-I\left({\rm A,E}\right)-I\left({\rm B,E}\right)+\left(\frac{P_{d}+{\rm QBER}}{2}\right)\right]\\
\\
P_{B}^{\mathcal{E}}(p,q) & = & 0.25\times\left[I\left({\rm A,B}\right)-I\left({\rm A,E}\right)-I\left({\rm B,E}\right)+\left(\frac{P_{d}+{\rm QBER}}{2}\right)\right]\\
\\
P_{E}^{\mathcal{E}}(p,q) & = & 0.25\times\left[-I\left({\rm A,B}\right)+I\left({\rm A,E}\right)+I\left({\rm B,E}\right)+\left(1-\frac{P_{d}+{\rm QBER}}{2}\right)\right]
\end{array}.\label{eq:Final_Payoffs}
\end{equation}
In this context, it is evident that the payoffs for Alice and Bob,
as denoted by Eq. (\ref{eq:Final_Payoffs}), are identical. For our
subsequent analysis, we will treat these individual payoffs independently.
This approach is taken due to the distinct probabilities, represented
by $q$ for Alice and $p$ for Bob, pertaining to the selection of
encoded bit values and the utilization of the $Z$ basis to prepare
initial states. This is illustrated more clearly
in the tabular format of the game $\mathcal{M}$ in Table \ref{tab:Payoff-matrix-Table}.

\begin{table}[h]
\begin{centering}
\begin{tabular}{|c|c|cc|cc|}
\hline 
 & \multicolumn{5}{c|}{Bob}\tabularnewline
\hline 
\multirow{8}{*}{Alice} &  & $Z\left(p\right)$ &  &  & $X\left(1-p\right)$\tabularnewline
\cline{2-6} \cline{3-6} \cline{4-6} \cline{5-6} \cline{6-6} 
 &  & $\left(rP_{A}^{E_{i}}\left(p,q\right)+\left(1-r\right)P_{A}^{E_{j}}\left(p,q\right),\right.$ &  &  & $\left(rP_{A}^{E_{i}}\left(1-p,q\right)+\left(1-r\right)P_{A}^{E_{j}}\left(1-p,q\right),rP_{B}^{E_{i}}\left(1-p,q\right)\right.$\tabularnewline
 & $0\left(q\right)$ & $\left.rP_{B}^{E_{i}}\left(p,q\right)+\left(1-r\right)P_{B}^{E_{j}}\left(p,q\right),P_{E}^{E_{i/j}}\left(p,q\right)\right)$ &  &  & $\left.+\left(1-r\right)P_{B}^{E_{j}}\left(1-p,q\right),P_{E}^{E_{i/j}}\left(1-p,q\right)\right)$\tabularnewline
 &  &  &  &  & \tabularnewline
\cline{2-6} \cline{3-6} \cline{4-6} \cline{5-6} \cline{6-6} 
 &  &  &  &  & \tabularnewline
 & $1\left(1-q\right)$ & $\left(rP_{A}^{E_{i}}\left(p,1-q\right)+\left(1-r\right)P_{A}^{E_{j}}\left(p,1-q\right),\right.$ &  &  & $\left(rP_{A}^{E_{i}}\left(1-p,1-q\right)+\left(1-r\right)P_{A}^{E_{j}}\left(1-p,1-q\right),\right.$\tabularnewline
 &  & $rP_{B}^{E_{i}}\left(p,1-q\right)+\left(1-r\right)P_{B}^{E_{j}}\left(p,1-q\right),$ &  &  & $rP_{B}^{E_{i}}\left(1-p,1-q\right)+\left(1-r\right)P_{B}^{E_{j}}\left(1-p,1-q\right),$\tabularnewline
 &  & $\left.P_{E}^{E_{i/j}}\left(p,1-q\right)\right)$ &  &  & $\left.P_{E}^{E_{i/j}}\left(1-p,1-q\right)\right)$\tabularnewline
\hline 
\end{tabular}
\par\end{centering}
\caption{\label{tab:Payoff-matrix-Table}Payoff matrix for
the $\mathcal{M}$ game. The first, second and third entry in the
parenthesis denotes the payoff of Alice, Bob and Eve. Here, $r$ and
$1-r$ denote the probabilities assigned to the selection of attacks
$E_{i}$ and $E_{j}$ by an eavesdropper, Eve, respectively.}

\end{table}

\section{Investigating Threshold Bound of QBER using Nash Equilibrium\label{sec:III}}

A Nash equilibrium is a set of actions in a game where no player can
improve their expected outcome (payoff) by altering their choice unilaterally,
assuming the other players' decisions remain unchanged. In a generalized
concept of Nash equilibrium that represents a stochastic steady state
of a strategic game, each player has the option to select a probability
distribution over their available actions instead of being limited
to a single, fixed choice. This probability distribution is referred
to as a mixed strategy. In a well-defined game, it is generally assumed
that all participants act logically and rationally. As a result, the
primary goal for all players is to optimize (maximize in our case
) their expected payoffs. For simplicity, we evaluate \emph{mixed
strategy Nash equilibrium}\footnote{A mixed strategy Nash equilibrium in a normal-form game is a set of
mixed strategies for each player, where no player has an incentive
to unilaterally deviate given the strategies chosen by others, ensuring
mutual optimality.} by taking three sets of game scenarios $E_{1}$-$E_{2}$,
$E_{1}$-$E_{3}$ and $E_{2}$-$E_{3}$ scenario where each player
is looking to play a mixed quantum strategy that makes her opponent
indifferent between her pure quantum strategies\footnote{All payoff elements for all parties under the attack scenarios $E_{1}$,
$E_{2}$, $E_{3}$ and $E_{4}$ are computed in Appendices A, B,
C and D, respectively.}. 
Based on this assumption, we can compare the results derived from
Nash equilibrium points to obtain the secure bound of QBER (see Appendix
E).

Firstly, we elucidate the best response function and subsequently
apply it to various game scenarios. For each game
scenario, we will establish
the best response function for each party. The best response for an
individual is defined as the probability of choosing their rational
choice in a mixed strategy scenario, aiming to achieve the best utility
when other parties make their decisions independently and arbitrarily.
Let's consider the best response function for Alice in the game
$E_{i}$ - $E_{j}$. Suppose Alice has the probability $q$ of choosing
her classical bit information when Bob and Eve independently choose
their decisions with probabilities $p$ and\footnote{Here, $r$ and $1-r$ denote the probabilities assigned to the selection
of attacks $E_{i}$ and $E_{j}$ by an eavesdropper, Eve, respectively.} $r$. For the pure strategy best response function of Alice, denoted
as $q=1$ and $q=0$, it can be expressed as \cite{HHM10}:

\[
B_{A}\left(p,r\right)\coloneqq rP_{A}^{E_{i}}\left(p,1\right)+\left(1-r\right)P_{A}^{E_{j}}\left(p,1\right)>rP_{A}^{E_{i}}\left(p,0\right)+\left(1-r\right)P_{A}^{E_{j}}\left(p,0\right).
\]
and

\[
B_{A}\left(p,r\right)\coloneqq rP_{A}^{E_{i}}\left(p,1\right)+\left(1-r\right)P_{A}^{E_{j}}\left(p,1\right)<rP_{A}^{E_{i}}\left(p,0\right)+\left(1-r\right)P_{A}^{E_{j}}\left(p,0\right),
\]
respectively. For the mixed strategy best response function of Alice,
is given by:

\[
B_{A}\left(p,r\right)\coloneqq rP_{A}^{E_{i}}\left(p,1\right)+\left(1-r\right)P_{A}^{E_{j}}\left(p,1\right)=rP_{A}^{E_{i}}\left(p,0\right)+\left(1-r\right)P_{A}^{E_{j}}\left(p,0\right).
\]
Using the definition of the best response function, we can similarly
define the best response functions for Bob and Eve as $B_{B}\left(q,r\right)$
and $B_{E}\left(p,q\right)$, where Bob and Eve assign their probabilities
$p$ and $r$ with best responses to $q,r$ and $p,q$, respectively.

\emph{$E_{1}$-$E_{2}$ }\emph{game scenario} Previously defined probabilities
$p$ and $q$ can be used. Let's assume that Eve chooses the $E_{1}$($E_{2}$)
attack with probability $r$ ($1-r$). We denote the \emph{best response
function} of Alice as $B_{A}\left(p,r\right)$, which represents the
set of probabilities that Alice assigns to choosing the Z basis in
her best responses to $p$ and $r$ \cite{HHM10}. Similarly, the best response
functions for Bob and Eve are $B_{B}\left(q,r\right)$ and $B_{E}\left(p,q\right)$,
respectively. We have,

\begin{equation}
\begin{array}{lcl}
B_{A}\left(p,r\right) & = & \begin{cases}
\left\{ q=1\right\}  & {\rm if}\,\,rP_{A}^{E_{1}}\left(p,1\right)+\left(1-r\right)P_{A}^{E_{2}}\left(p,1\right)>rP_{A}^{E_{1}}\left(p,0\right)+\left(1-r\right)P_{A}^{E_{2}}\left(p,0\right)\\
\left\{ q:0\le q\le1\right\}  & {\rm if}\,\,rP_{A}^{E_{1}}\left(p,1\right)+\left(1-r\right)P_{A}^{E_{2}}\left(p,1\right)=rP_{A}^{E_{1}}\left(p,0\right)+\left(1-r\right)P_{A}^{E_{2}}\left(p,0\right)\\
\left\{ q=0\right\}  & {\rm if}\,\,rP_{A}^{E_{1}}\left(p,1\right)+\left(1-r\right)P_{A}^{E_{2}}\left(p,1\right)<rP_{A}^{E_{1}}\left(p,0\right)+\left(1-r\right)P_{A}^{E_{2}}\left(p,0\right)
\end{cases}\\
\\
B_{B}\left(q,r\right) & = & \begin{cases}
\left\{ p=1\right\}  & {\rm if}\,\,rP_{B}^{E_{1}}\left(1,q\right)+\left(1-r\right)P_{B}^{E_{2}}\left(1,q\right)>rP_{B}^{E_{1}}\left(0,q\right)+\left(1-r\right)P_{B}^{E_{2}}\left(0,q\right)\\
\left\{ p:0\le p\le1\right\}  & {\rm if}\,\,rP_{B}^{E_{1}}\left(1,q\right)+\left(1-r\right)P_{B}^{E_{2}}\left(1,q\right)=rP_{B}^{E_{1}}\left(0,q\right)+\left(1-r\right)P_{B}^{E_{2}}\left(0,q\right)\\
\left\{ p=0\right\}  & {\rm if}\,\,rP_{B}^{E_{1}}\left(1,q\right)+\left(1-r\right)P_{B}^{E_{2}}\left(1,q\right)<rP_{B}^{E_{1}}\left(0,q\right)+\left(1-r\right)P_{B}^{E_{2}}\left(0,q\right)
\end{cases}\\
\\
B_{E}\left(p,q\right) & = & \begin{cases}
\left\{ r=1\right\}  & {\rm if}\,\,P_{E}^{E_{1}}\left(p,q\right)>P_{E}^{E_{2}}\left(p,q\right)\\
\left\{ r:0\le r\le1\right\}  & {\rm if}\,\,P_{E}^{E_{1}}\left(p,q\right)=P_{E}^{E_{2}}\left(p,q\right)\\
\left\{ r=0\right\}  & {\rm if}\,\,P_{E}^{E_{1}}\left(p,q\right)<P_{E}^{E_{2}}\left(p,q\right)
\end{cases}
\end{array}.\label{eq:Best_Response_Function_E1_E2_Attack}
\end{equation}
The best response functions for Alice, Bob and Eve in the game
scenarios \emph{$E_{1}$-$E_{3}$} and \emph{$E_{2}$-$E_{3}$} are
as follows,

\begin{equation}
\begin{array}{lcl}
B_{A}\left(p,r\right) & = & \begin{cases}
\left\{ q=1\right\}  & {\rm if}\,\,rP_{A}^{E_{1}}\left(p,1\right)+\left(1-r\right)P_{A}^{E_{3}}\left(p,1\right)>rP_{A}^{E_{1}}\left(p,0\right)+\left(1-r\right)P_{A}^{E_{3}}\left(p,0\right)\\
\left\{ q:0\le q\le1\right\}  & {\rm if}\,\,rP_{A}^{E_{1}}\left(p,1\right)+\left(1-r\right)P_{A}^{E_{3}}\left(p,1\right)=rP_{A}^{E_{1}}\left(p,0\right)+\left(1-r\right)P_{A}^{E_{3}}\left(p,0\right)\\
\left\{ q=0\right\}  & {\rm if}\,\,rP_{A}^{E_{1}}\left(p,1\right)+\left(1-r\right)P_{A}^{E_{3}}\left(p,1\right)<rP_{A}^{E_{1}}\left(p,0\right)+\left(1-r\right)P_{A}^{E_{3}}\left(p,0\right)
\end{cases}\\
\\
B_{B}\left(q,r\right) & = & \begin{cases}
\left\{ p=1\right\}  & {\rm if}\,\,rP_{B}^{E_{1}}\left(1,q\right)+\left(1-r\right)P_{B}^{E_{3}}\left(1,q\right)>rP_{B}^{E_{1}}\left(0,q\right)+\left(1-r\right)P_{B}^{E_{3}}\left(0,q\right)\\
\left\{ p:0\le p\le1\right\}  & {\rm if}\,\,rP_{B}^{E_{1}}\left(1,q\right)+\left(1-r\right)P_{B}^{E_{3}}\left(1,q\right)=rP_{B}^{E_{1}}\left(0,q\right)+\left(1-r\right)P_{B}^{E_{3}}\left(0,q\right)\\
\left\{ p=0\right\}  & {\rm if}\,\,rP_{B}^{E_{1}}\left(1,q\right)+\left(1-r\right)P_{B}^{E_{3}}\left(1,q\right)<rP_{B}^{E_{1}}\left(0,q\right)+\left(1-r\right)P_{B}^{E_{3}}\left(0,q\right)
\end{cases}\\
\\
B_{E}\left(p,q\right) & = & \begin{cases}
\left\{ r=1\right\}  & {\rm if}\,\,P_{E}^{E_{1}}\left(p,q\right)>P_{E}^{E_{3}}\left(p,q\right)\\
\left\{ r:0\le r\le1\right\}  & {\rm if}\,\,P_{E}^{E_{1}}\left(p,q\right)=P_{E}^{E_{3}}\left(p,q\right)\\
\left\{ r=0\right\}  & {\rm if}\,\,P_{E}^{E_{1}}\left(p,q\right)<P_{E}^{E_{3}}\left(p,q\right)
\end{cases}
\end{array}.\label{eq:Best_Response_Function_E1_E2_Attack-1}
\end{equation}
and

\begin{equation}
\begin{array}{lcl}
B_{A}\left(p,r\right) & = & \begin{cases}
\left\{ q=1\right\}  & {\rm if}\,\,rP_{A}^{E_{2}}\left(p,1\right)+\left(1-r\right)P_{A}^{E_{3}}\left(p,1\right)>rP_{A}^{E_{2}}\left(p,0\right)+\left(1-r\right)P_{A}^{E_{3}}\left(p,0\right)\\
\left\{ q:0\le q\le1\right\}  & {\rm if}\,\,rP_{A}^{E_{2}}\left(p,1\right)+\left(1-r\right)P_{A}^{E_{3}}\left(p,1\right)=rP_{A}^{E_{2}}\left(p,0\right)+\left(1-r\right)P_{A}^{E_{3}}\left(p,0\right)\\
\left\{ q=0\right\}  & {\rm if}\,\,rP_{A}^{E_{2}}\left(p,1\right)+\left(1-r\right)P_{A}^{E_{3}}\left(p,1\right)<rP_{A}^{E_{2}}\left(p,0\right)+\left(1-r\right)P_{A}^{E_{3}}\left(p,0\right)
\end{cases}\\
\\
B_{B}\left(q,r\right) & = & \begin{cases}
\left\{ p=1\right\}  & {\rm if}\,\,rP_{B}^{E_{2}}\left(1,q\right)+\left(1-r\right)P_{B}^{E_{3}}\left(1,q\right)>rP_{B}^{E_{2}}\left(0,q\right)+\left(1-r\right)P_{B}^{E_{3}}\left(0,q\right)\\
\left\{ p:0\le p\le1\right\}  & {\rm if}\,\,rP_{B}^{E_{2}}\left(1,q\right)+\left(1-r\right)P_{B}^{E_{3}}\left(1,q\right)=rP_{B}^{E_{2}}\left(0,q\right)+\left(1-r\right)P_{B}^{E_{3}}\left(0,q\right)\\
\left\{ p=0\right\}  & {\rm if}\,\,rP_{B}^{E_{2}}\left(1,q\right)+\left(1-r\right)P_{B}^{E_{3}}\left(1,q\right)<rP_{B}^{E_{2}}\left(0,q\right)+\left(1-r\right)P_{B}^{E_{3}}\left(0,q\right)
\end{cases}\\
\\
B_{E}\left(p,q\right) & = & \begin{cases}
\left\{ r=1\right\}  & {\rm if}\,\,P_{E}^{E_{2}}\left(p,q\right)>P_{E}^{E_{3}}\left(p,q\right)\\
\left\{ r:0\le r\le1\right\}  & {\rm if}\,\,P_{E}^{E_{2}}\left(p,q\right)=P_{E}^{E_{3}}\left(p,q\right)\\
\left\{ r=0\right\}  & {\rm if}\,\,P_{E}^{E_{2}}\left(p,q\right)<P_{E}^{E_{3}}\left(p,q\right)
\end{cases}
\end{array}.\label{eq:Best_Response_Function_E1_E2_Attack-2}
\end{equation}

\begin{figure}[h]
\centering{}\includegraphics[scale=0.4]{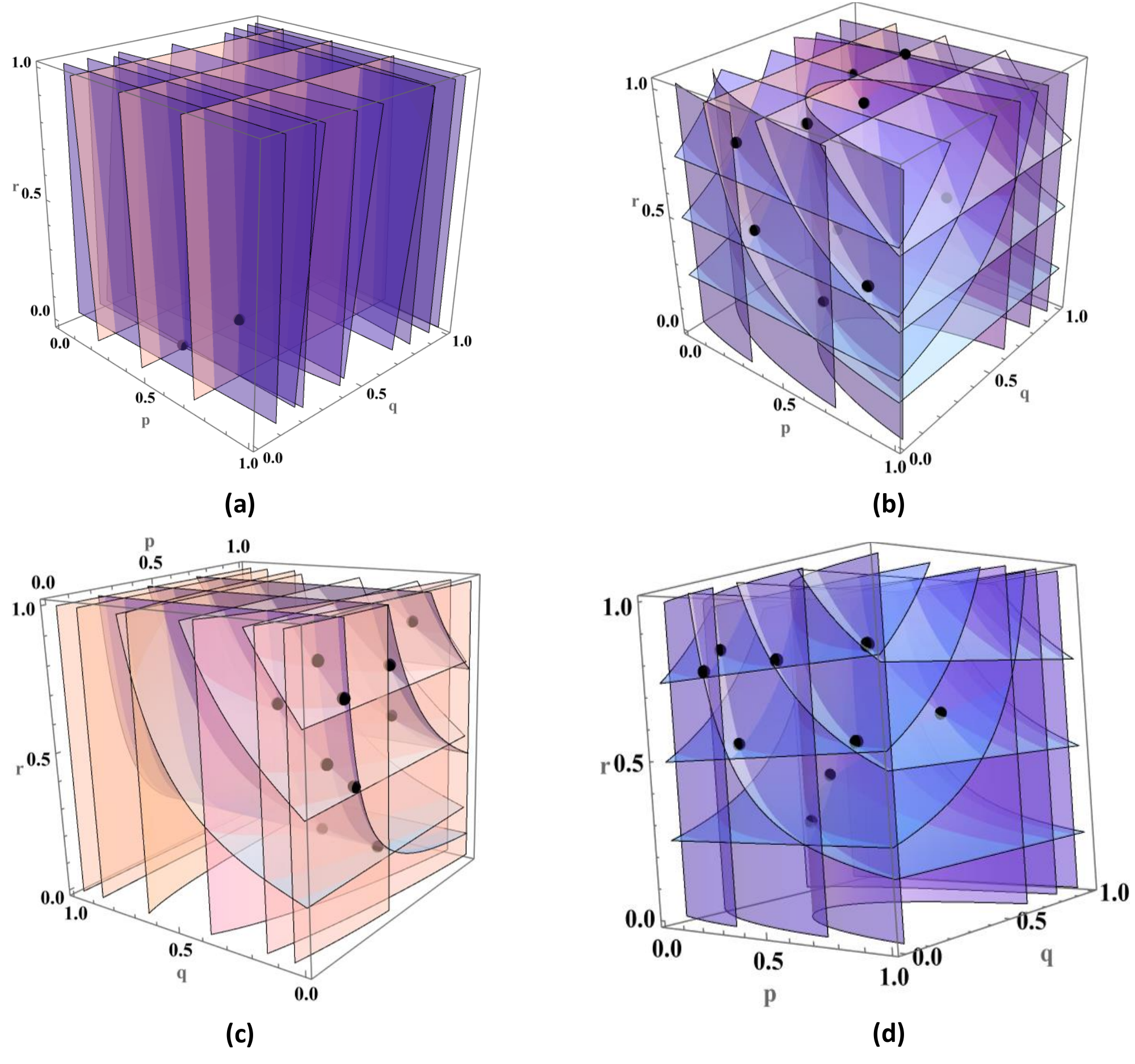}\caption{\label{fig:Nash_Equilibrium}(Color online) The plot shows Nash equilibrium points
where the best response functions of the three players intersect in
our mixed strategy game scenario. The low-density layer, medium-density
layer, and high-density layer correspond to the best response functions
of Alice, Bob, and Eve, respectively: (a) $E_{1}$-$E_{2}$ game scenario,
(b) $E_{1}$-$E_{3}$ game scenario, (c) $E_{2}$-$E_{3}$ game scenario, (d)
$E_{1}$-$E_{4}$ game scenario.}
\end{figure}

 In the context
of a quantum game, Figure \ref{fig:Nash_Equilibrium} shows the best
response functions for different game
scenarios. These functions represent
each player's optimal strategy in response to the strategies of the
other players. The mixed strategy Nash equilibrium is found at the
points where these best response functions intersect. This means that
at these intersections, each player's strategy is optimal given the
strategies of the others. It is important to note that there can be
multiple intersection points, each corresponding to different game scenarios. These points of intersection indicate the Nash equilibrium
for each player in these game scenarios. In subfigures (a), (b), (c) and
(d) of Figure \ref{fig:Nash_Equilibrium}, we identify the Nash equilibrium
points for the $E_{1}$-$E_{2}$, $E_{1}$-$E_{3}$, $E_{2}$-$E_{3}$
and $E_{1}$-$E_{4}$ game scenarios, respectively. Each subfigure depicts
three different curves, each representing the best response function
of a player within the mixed quantum strategy game. These curves show
the optimal strategies for each player in response to the others.
The intersection points of these three curves indicate the most stable
points in each game
scenario. From these intersections, we can determine
the most stable probability distributions in the mixed quantum strategy
game, which are optimal for all players.

 We provide a summary of this data, including the payoffs of each player, in Table
\ref{tab:NE_Points_Payoff} in Appendix E. In the provided summary,
we analyze our mixed strategy Nash equilibrium points for different
game scenarios and investigate the secure QBER bound. In the context of
Eq. (\ref{eq:Final_Payoffs}), it becomes evident that the payoff
functions for the three parties depend on the variables $p,q$ ($q$)
for Eve's attack strategies, $E_{1}$, $E_{2}$ ( $E_{3}$, $E_{4}$).
Furthermore, our game can be represented as a normal form game (strategic
form game) because players have no information at their decision points
about other players' choices when making their moves. In this setup,
all parties independently select their strategies. Specifically, Bob
selects the $Z$ basis with a probability of $p$, Eve chooses the
attack strategy with a probability of $r$, and Alice, in message
mode, selects the encoded bit $0$ with a probability of $q$. In
our analysis, both Alice and Bob share the same payoff functions (\ref{eq:Final_Payoffs})).
In the $E_{1}$-$E_{2}$ game scenario, the expected payoffs for Alice/Bob
and Eve are given by $rP_{A/B}^{E_{1}}(p,q)+(1-r)P_{A/B}^{E_{2}}(p,q)$
and $rP_{E}^{E_{1}}(p,q)+(1-r)P_{E}^{E_{2}}(p,q)$, respectively.
By comparing the payoff differences between Eve and Alice (see in
Table \ref{tab:NE_Points_Payoff}), we can identify among the Nash
equilibrium points where Eve or Alice/Bob benefit the most. In the
$E_{1}$-$E_{2}$  game scenario, Eve benefits individually at the Nash equilibrium
point $(0.45,0.195,0.005)$, while Alice/Bob benefits individually
at the Nash equilibrium point $(0.72,0.208,0.225)$. Similarly, in
the $E_{1}$-$E_{3}$ game scenario, Eve and Alice/Bob benefit individually
at Nash equilibrium points $\left(0.41,0.39,0.412\right)$ and $\left(0.84,0.047,0.525\right)$,
respectively. In the $E_{2}$-$E_{3}$ game scenario, Eve and Alice/Bob
benefit individually at Nash equilibrium points $\left(0.385,0.215,0.262\right)$
and $\left(0.80,0.115,0.885\right)$, respectively. In conclusion,
there is no Pareto optimal Nash equilibrium point in our game scenarios
where payoffs would be favorable to the entire group of players.

We calculate the QBER for all Nash equilibrium points in different
game scenarios. It is pertinent to highlight that the expected QBER value
for the $E_{i}$-$E_{j}$ game scenario is defined as $\epsilon_{E_{i}-E_{2}}=r\,{\rm QBER}_{E_{i}}+\left(1-r\right){\rm QBER}_{E_{j}}$,
where $r$ is a probability of Eve's choice to perform $E_{i}$ attack.
In any given game scenario, Eve's optimal scenario is characterized by either
the maximum payoff difference (the discrepancy between Eve's and Alice's
payoffs) or the minimum QBER. Our objective is to determine the QBER
threshold value. Consequently, we seek the minimum QBER values across
all game  scenarios outlined in Table \ref{tab:NE_Points_Payoff}.
These values correspond to the Nash equilibrium points, representing
situations where Eve is strategically positioned to conceal her presence
most effectively. The minimum QBER values at Nash equilibrium points
are $0.610303$, $0.152451$ and $0.143882$ for the $E_{1}$-$E_{2}$,
$E_{1}$-$E_{3}$ and $E_{2}$-$E_{3}$ game scenarios, respectively. The
reduction in the minimum QBER value\footnote{Considering lower QBER value empower Eve the most, that bound gives
the most secure condition on a quantum protocol.} (in message mode) suggests that more potent attack strategies by
Eve are being applied in this game scenario when the minimum or same $P_{d}$
(detection probability of Eve's presence in control mode) is achieved
for the attacks within this
game scenario. These lower QBER values are designed
to identify more sophisticated quantum attacks by Eve within the game-like
scenario. Following an analysis of the threshold values of QBER, we
conclude that $E_{3}$ represents the most powerful quantum attack,
followed by $E_{2}$, and then $E_{1}$. This implies that whenever
Eve employs a powerful attack, she attains a higher payoff, posing
a threat to both Alice and Bob. To address this threat in our DL04
protocol game scenario, we set a lower minimum value for the QBER
corresponding to this specific potent attack by Eve. The threshold
is determined by evaluating all Nash equilibrium points within these
game scenarios that encompass the mentioned attack.

As previously noted, a player with limited access to quantum resources
appears to be a classical player. In our analysis, the $E_{4}$ attack
is considered a classical attack by Eve, implying she is a classical
eavesdropper when performing the $E_{4}$ attack. We also compare
the classical attack $E_{4}$ with the simplest quantum attack, $E_{1}$.
Furthermore, we analyze the $E_{4}$ attack (IR attack) in comparison
with the $E_{1}$ attack as a $E_{1}$-$E_{4}$ game
scenario. In this game scenario,
the minimum value of QBER is found to be $0.323478$, which is lower
than $E_{1}$-$E_{2}$ game scenario. Despite this fact, $E_{4}$ is considered
a less powerful attack than $E_{1}$ because the $P_{d}$ value is
higher for $E_{4}$ (37.5\%) compared to that for $E_{1}$ (18.75\%)
in control mode. 

To determine the secure upper bound of the QBER, denoted as $\epsilon$,
considering all attack scenarios in the entire game, we focus on this
game scenario involving potent attacks which is the $E_{2}$-$E_{3}$ game scenario.
In this context, the upper bound of QBER is found to be $\epsilon=0.143882$
because this is the minimum QBER value for the game scenario where
the most potent attack $E_{3}$ is present. The lower bound of QBER
is $0$ because if Eve applies only the $E_{3}$ attack, no error
will be detected in message mode (see Appendix C). Furthermore, the
lower and upper bounds of the detection probability of Eve's presence
($P_{d}$) in control mode are established at $0.1875$ and $0.375$,
respectively. Consequently, we can deduce that the DL04 protocol demonstrates
$\epsilon$-secure, under the set of collective and individual (IR)
attacks, denoted as $E_{1}$, $E_{2}$, $E_{3}$ and $E_{4}$.

\section{Discussion\label{sec:IV}}

In this paper, we propose a new security definition for quantum communication
protocols in the context of collective attacks and IR attack using
Nash equilibrium. Alternatively,
this can be viewed as a game-theoretic security bound against collective
attacks using Nash equilibrium. Nash equilibrium points are stable
points that provide rational probabilities for decision-making in
a mixed quantum strategy game. These probabilities create a stable
game situation for all parties, enabling the evaluation of cryptographic
parameters such as the threshold value of QBER. In a real-world scenario where decisions
are made independently by all parties, achieving a stable point is
less likely and may lead to a more unstable situation. Therefore,
the identification of the lowest QBER value from the stable game scenario
establishes a secure threshold boundary for the realistic implementation
of the quantum protocol. Our analysis illustrates that the security
(denoted as $\epsilon$-secure) of any quantum communication protocol
depends on the choices of attack strategies employed by the eavesdropper,
Eve. A smaller value of $\epsilon$ indicates a more potent attack
strategy by Eve. Furthermore, our findings suggest that a quantum
Eve possesses greater power than a classical Eve\footnote{Ensuring that Eve's selection of unitary operation and the execution
of the measurement operation are appropriately synchronized within
the quantum operation.}. In our investigation, we assume that Eve has unlimited quantum resources,
leading us to neglect the last three terms of Eve's payoff in Eq.
(\ref{eq:Initial_Payoffs}). To simplify our analysis, we also assume
equal contributions of all payoff elements for each party, with weighted
values set at $\frac{1}{4}$. We divide our game into four different
game scenarios
and determine the Nash equilibrium points for each game scenario. Subsequently,
we identify the minimum value of QBER within each game scenario, which corresponds
to the secure bound of QBER for that specific game scenario. Finally, we
determine the threshold bound of QBER ($\epsilon$) for the entire
game by evaluating the minimum QBER value within the game scenario that
encompasses the most potent attacks. Moreover, it has been observed
that the DL04 protocol is vulnerable to Pavi{\v{c}}i{\'c} attacks
in message mode due to the QBER being $0$. In this scenario, security
is ensured through the control mode executed jointly by both legitimate
parties. However, the security concern for this attack will be mitigated
through the control mode.

It is noteworthy that quantum communication protocols serve various
purposes, with security sometimes prioritized over efficiency, and
vice versa. To address such trade-off scenarios, one may generalize
our analysis by introducing $\omega_{i},\,\omega_{j},\,\omega_{k}\neq0$,
and $\omega_{m}\neq\omega_{n}$ if $m\neq n$. This implies distinct
weighted values with each payoff, with the sum normalized to $1$
(ensuring weights assigned to players are normalized). Assuming $\omega_{i},\,\omega_{j},\,\omega_{k}\neq0$,
Eq. (\ref{eq:Initial_Payoffs}) yields $n_{1}=1$, $n_{2}=1$, $n_{3}=1$
for $E_{1}$, $n_{1}=4$, $n_{2}=2$, $n_{3}=1$ for $E_{2}$, $n_{1}=2$,
$n_{2}=5$, $n_{3}=0$ for $E_{3}$ and $n_{1}=2$, $n_{2}=n_{3}=0$
for $E_{4}$ attack strategies. It is important to
note that the cost of using multi-qubit gates goes up as the number
of qubits in the gate increases. Importantly, our approach can be further
generalized to encompass an entire game-like scenario with all these
attacks together rather than a set of different game scenarios. Under this assumption,
different probabilities would exist for applying various attacks by
eavesdroppers, leading to the identification of Pareto optimal Nash
equilibrium points under specific conditions. Consequently, the expression
for the QBER bound ($\epsilon$) becomes dependent on these generalized
parameters.

Our methodology for establishing the securely bounded threshold of
QBER is employed in the DL04 protocol. This approach is adaptable
to both existing and future protocols, allowing for the attainment
of diverse information theoretic bounds. This flexibility accommodates
various strategies implemented by stakeholders and aligns with the
purpose of applying quantum protocols. We defer further investigation
to integrate additional game-theoretic features suitable for various
purposes in different quantum protocols, considering other attributes
of quantum mechanics. This collaborative effort aims to yield more
robust results, enhancing the analysis of quantum information bounds.

\subsection*{Acknowledgment: }

Authors acknowledge support from the QUEST scheme of the Interdisciplinary
Cyber-Physical Systems (ICPS) program of the Department of Science
and Technology (DST), India, Grant No.: DST/ICPS/QuST/Theme-1/2019/14
(Q80). They also thank R. Srikanth for his interest and useful technical
feedback on this work.

\section*{Availability of data and materials}

No additional data is needed for this work.

\section*{Competing interests}

The authors declare that they have no competing interests.

\bibliographystyle{apsrev4-2}
\bibliography{QuantumGame}

\appendix
%dummy comment inserted by tex2lyx to ensure that this paragraph is not empty

\section*{Appendix A }

We employ W{\'o}jcik's attack strategy \cite{W03} as $E_{1}$ targeting
the DL04 protocol in both message and control modes. As previously
mentioned, Bob prepares the initial state randomly, choosing from
$|0\rangle$, $|1\rangle$, $|+\rangle$ and $|-\rangle$ with probabilities
of $\frac{p}{2}$, $\frac{p}{2}$, $\frac{1-p}{2}$ and $\frac{1-p}{2}$,
respectively. Alice encodes $j$ where $j\in\{0,1\}$ onto the travel
photon\footnote{Here, $t$ represents the travel photon (qubit) by Bob (Alice) for
B-A attack (A-B attack).} (indexed with the subscript $t$) received from Bob using operation\footnote{Alice performs operation $iY_{t}^{0}\equiv I_{t}$ $\left(iY_{t}^{1}\equiv\left(ZX\right)_{t}^{1}\right)$
to encode $0$ $\left(1\right)$ with probability $q$ $\left(1-q\right)$.} $iY_{t}^{j}$. We assume that Alice and Bob utilize a lossy quantum
channel with single-photon transmission efficiency denoted as $\eta$.
In contrast, Eve replaces the channel with a perfect one with $\eta=1$.
Eve employs two auxiliary spatial modes, $x$ and $y$, along with
a single photon initially in the state $|0\rangle$. She conducts
eavesdropping on the quantum channel twice: first during transmission
from Bob to Alice (B-A attack) and then during transmission from Alice
to Bob (A-B attack). The eavesdropping protocol (refer to Fig. 2 in
Ref. \cite{W03}) begins by preparing two auxiliary modes, $x$ and
$y$, in the state $|{\rm vac}\rangle_{x}|0\rangle_{y}$, where $|{\rm vac}\rangle$
represents an empty mode. Eve uses the unitary operation $Q_{txy}$
during the B-A attack and its conjugate, $Q_{txy}^{\dagger}\left(\equiv Q_{txy}^{-1}\right)$,
during the A-B attack. The unitary operator $Q_{txy}(={\rm SWAP}_{tx}\,{\rm CPBS}_{txy}\,H_{y}\equiv{\rm SWAP}_{tx}\otimes I_{y}\,{\rm CPBS}_{txy}\,I_{t}\otimes I_{x}\otimes H_{y})$\textcolor{red}{{}
}is composed of a Hadamard gate (a one-qubit gate), a SWAP gate (a
two-qubit gate) and a controlled polarizing beam splitter (CPBS),
which is a three-qubit gate\footnote{polarizing beam splitter which is assumed to transmit (reflect) photons
in the state $|0\rangle$ $\left(|1\rangle\right)$.}. We will now analyze this attack strategy for both message mode and
control mode to derive the components of the payoff functions for
Alice, Bob and Eve.

\textbf{\emph{Message mode:}} 
To begin with simplicity, let us focus on scenarios where Alice encodes
0. Specifically, we will examine the situation where Bob prepares
the initial state $|0\rangle_{t}$. We will then discuss Eve's attack
strategy on the composite system, which involves the state $|0\rangle_{t}|{\rm vac}\rangle_{x}|0\rangle_{y}$
and is denoted as $E_{1}$,

\[
\begin{array}{lcl}
|{\rm B-A}\rangle_{|0\rangle E_{1}} & = & Q_{txy}\,|0\rangle_{t}|{\rm vac}\rangle_{x}|0\rangle_{y}\\
 & = & {\rm SWAP}_{tx}\,{\rm CPBS}_{txy}\,H_{y}\,\left(|0\rangle|{\rm vac}\rangle|0\rangle\right)_{txy}\\
 & = & {\rm SWAP}_{tx}\,{\rm CPBS}_{txy}\,\frac{1}{\sqrt{2}}\left(|0\rangle|{\rm vac}\rangle|0\rangle+|0\rangle|{\rm vac}\rangle|1\rangle\right)_{txy}\\
 & = & {\rm SWAP}_{tx}\,\frac{1}{\sqrt{2}}\left(|0\rangle|0\rangle|{\rm vac}\rangle+|0\rangle|{\rm vac}\rangle|1\rangle\right)_{txy}\\
 & = & \frac{1}{\sqrt{2}}\left[|0\rangle|0\rangle|{\rm vac}\rangle+|{\rm vac}\rangle|0\rangle|1\rangle\right]_{txy}
\end{array},
\]
as Alice performs the operation $iY_{t}^{0}$ (to encode $0$) on
$t$ state, the composite system $|{\rm B-A}\rangle_{|0\rangle E_{1}}$will
remain the same and will be symbolically denoted as $|{\rm B-A}\rangle_{|0\rangle E_{1}}^{0}$with
the superscript denoting the bit encoded by Alice. The modified system
after Eve's A-B attack $\left(Q_{txy}^{-1}\right)$ is,

\begin{equation}
\begin{array}{lcl}
|{\rm A-B}\rangle_{|0\rangle E_{1}}^{0} & = & Q_{txy}^{-1}\,|{\rm B-A}\rangle_{|0\rangle E_{1}}^{0}\\
 & = & Q_{txy}^{-1}\,\frac{1}{\sqrt{2}}\left[|0\rangle|0\rangle|{\rm vac}\rangle+|{\rm vac}\rangle|0\rangle|1\rangle\right]_{txy}\\
 & = & H_{y}\,{\rm CPBS}_{txy}\,{\rm SWAP}_{tx}\,\frac{1}{\sqrt{2}}\left[|0\rangle|0\rangle|{\rm vac}\rangle+|{\rm vac}\rangle|0\rangle|1\rangle\right]_{txy}\\
 & = & H_{y}\,{\rm CPBS}_{txy}\,\frac{1}{\sqrt{2}}\left[|0\rangle|0\rangle|{\rm vac}\rangle+|0\rangle|{\rm vac}\rangle|1\rangle\right]_{txy}\\
 & = & H_{y}\,\frac{1}{\sqrt{2}}\left[|0\rangle|{\rm vac}\rangle|0\rangle+|0\rangle|{\rm vac}\rangle|1\rangle\right]_{txy}\\
 & = & |0\rangle_{t}|{\rm vac}\rangle_{x}|0\rangle_{y}
\end{array}.\label{eq:Final_State_Bob_Zero_Alice_Zero_E1}
\end{equation}

This outcome aligns with the initial composite system, as anticipated
because Alice employs an identity operation to encode $0$, while
Eve conducts a unitary A-B attack operation $\left(Q_{txy}\right)$
and its inverse, the unitary B-A attack operation $\left(Q_{txy}^{-1}\right)$.
Likewise, we can describe the remaining composite systems in which
Alice encodes $0$ when Bob prepares initial states $|1\rangle_{t}$,
$|+\rangle_{t}$ and $|-\rangle_{t}$, viz.,

\begin{equation}
\begin{array}{lcl}
|{\rm A-B}\rangle_{|1\rangle E_{1}}^{0} & = & |1\rangle_{t}|{\rm vac}\rangle_{x}|0\rangle_{y}\end{array},\label{eq:Final_State_Bob_One_Alice_Zero_E1}
\end{equation}

\begin{equation}
\begin{array}{lcl}
|{\rm A-B}\rangle_{|+\rangle E_{1}}^{0} & = & |+\rangle_{t}|{\rm vac}\rangle_{x}|0\rangle_{y}\end{array},\label{eq:Final_State_Bob_Plus_Alice_Zero_E1}
\end{equation}
and

\begin{equation}
\begin{array}{lcl}
|{\rm A-B}\rangle_{|-\rangle E_{1}}^{0} & = & |-\rangle_{t}|{\rm vac}\rangle_{x}|0\rangle_{y}\end{array},\label{eq:Final_State_Bob_Minus_Alice_Zero_E1}
\end{equation}
respectively.

Now, we consider the scenarios in which Alice encodes $1$ through
the operation $iY_{t}^{1}$ on the $t$ state. First, we consider
the system in which Bob's initial state is $|0\rangle_{t}$. We have
previously established that $|{\rm B-A}\rangle_{|0\rangle}=\frac{1}{\sqrt{2}}\left[|0\rangle|0\rangle|{\rm vac}\rangle+|{\rm vac}\rangle|0\rangle|1\rangle\right]_{txy}$.
Next, Alice performs her encoding operation, and then Eve executes
$Q_{txy}^{-1}$. The composite system is as follows,

\begin{equation}
\begin{array}{lcl}
|{\rm A-B}\rangle_{|0\rangle E_{1}}^{1} & = & Q_{txy}^{-1}\,iY_{t}^{1}\,\frac{1}{\sqrt{2}}\left[|0\rangle|0\rangle|{\rm vac}\rangle+|{\rm vac}\rangle|0\rangle|1\rangle\right]_{txy}\\
 & = & Q_{txy}^{-1}\,\frac{1}{\sqrt{2}}\left[-|1\rangle|0\rangle|{\rm vac}\rangle+|{\rm vac}\rangle|0\rangle|1\rangle\right]_{txy}\\
 & = & H_{y}\,{\rm CPBS}_{txy}\,{\rm SWAP}_{tx}\,\frac{1}{\sqrt{2}}\left[-|1\rangle|0\rangle|{\rm vac}\rangle+|{\rm vac}\rangle|0\rangle|1\rangle\right]_{txy}\\
 & = & H_{y}\,{\rm CPBS}_{txy}\,\frac{1}{\sqrt{2}}\left[-|0\rangle|1\rangle|{\rm vac}\rangle+|0\rangle|{\rm vac}\rangle|1\rangle\right]_{txy}\\
 & = & H_{y}\,\frac{1}{\sqrt{2}}\left[-|0\rangle|1\rangle|{\rm vac}\rangle+|0\rangle|{\rm vac}\rangle|1\rangle\right]_{txy}\\
 & = & \left[-\frac{1}{\sqrt{2}}|0\rangle|1\rangle|{\rm vac}\rangle+\frac{1}{2}|0\rangle|{\rm vac}\rangle|0\rangle-\frac{1}{2}|0\rangle|{\rm vac}\rangle|1\rangle\right]_{txy}
\end{array}.\label{eq:Final_State_Bob_Zero_Alice_One}
\end{equation}
In the interest of conciseness, we will not explicitly detail similar
calculations. From this point onward, we will only highlight the main
results. Let us now consider the scenario where Bob's initial state
$|1\rangle_{t}$ transforms into the composite system after the B-A
attack

\[
\begin{array}{lcl}
|{\rm B-A}\rangle_{|1\rangle E_{1}} & = & Q_{txy}\,|1\rangle_{t}|{\rm vac}\rangle_{x}|0\rangle_{y}\\
 & = & \frac{1}{\sqrt{2}}\left[|{\rm vac}\rangle|1\rangle|0\rangle\rangle+|1\rangle|1\rangle|{\rm vac}\rangle\right]_{txy}
\end{array}.
\]
The composite system resulting from Alice's operation and Eve's A-B
attack is,

\begin{equation}
\begin{array}{lcl}
|{\rm A-B}\rangle_{|1\rangle E_{1}}^{1} & = & Q_{txy}^{-1}\,iY_{t}^{1}\,\frac{1}{\sqrt{2}}\left[|{\rm vac}\rangle|1\rangle|0\rangle\rangle+|1\rangle|1\rangle|{\rm vac}\rangle\right]_{txy}\\
 & = & \left[\frac{1}{\sqrt{2}}|1\rangle|0\rangle|{\rm vac}\rangle+\frac{1}{2}|1\rangle|{\rm vac}\rangle|0\rangle+\frac{1}{2}|1\rangle|{\rm vac}\rangle|1\rangle\right]_{txy}
\end{array}.\label{eq:Final_State_Bob_One_Alice_One}
\end{equation}
The additional results for Bob's initial states $|+\rangle$ and $|-\rangle$
are,

\[
\begin{array}{lcl}
|{\rm B-A}\rangle_{|+\rangle E_{1}} & = & \frac{1}{2}\left[|0\rangle|0\rangle|{\rm vac}\rangle+|{\rm vac}\rangle|0\rangle|1\rangle+|{\rm vac}\rangle|1\rangle|0\rangle+|1\rangle|1\rangle|{\rm vac}\rangle\right]_{txy}\end{array},
\]

\begin{equation}
\begin{array}{lcl}
|{\rm A-B}\rangle_{|+\rangle E_{1}}^{1} & = & \left[\frac{1}{2}\left\{ |+\rangle|{\rm vac}\rangle|0\rangle-|-\rangle|{\rm vac}\rangle|1\rangle\right\} +\frac{1}{2\sqrt{2}}\left\{ -|+\rangle|1\rangle|{\rm vac}\rangle-|-\rangle|1\rangle|{\rm vac}\rangle+|+\rangle|0\rangle|{\rm vac}\rangle-|-\rangle|0\rangle|{\rm vac}\rangle\right\} \right]_{txy}\end{array},\label{eq:Final_State_Bob_Plus_Alice_One}
\end{equation}
and

\[
\begin{array}{lcl}
|{\rm B-A}\rangle_{|-\rangle E_{1}} & = & \frac{1}{2}\left[|0\rangle|0\rangle|{\rm vac}\rangle+|{\rm vac}\rangle|0\rangle|1\rangle-|{\rm vac}\rangle|1\rangle|0\rangle-|1\rangle|1\rangle|{\rm vac}\rangle\right]_{txy}\end{array},
\]

\begin{equation}
\begin{array}{lcl}
|{\rm A-B}\rangle_{|-\rangle E_{1}}^{1} & = & \left[\frac{1}{2}\left\{ |-\rangle|{\rm vac}\rangle|0\rangle-|+\rangle|{\rm vac}\rangle|1\rangle\right\} +\frac{1}{2\sqrt{2}}\left\{ -|+\rangle|1\rangle|{\rm vac}\rangle-|-\rangle|1\rangle|{\rm vac}\rangle-|+\rangle|0\rangle|{\rm vac}\rangle+|-\rangle|0\rangle|{\rm vac}\rangle\right\} \right]_{txy}\end{array}.\label{eq:Final_State_Bob_Minus_Alice_One}
\end{equation}
In the main text, we previously discussed that Bob can decode Alice's
message ($j$) deterministically by measuring the qubit on the same
basis he initially prepared it, without the need for a classical announcement.
The information obtained by Bob after decoding is represented as $m$.
Now, based on the results mentioned above, we can straightforwardly
identify the optimal strategy for Eve to infer Alice's encoded information
from her measurement results on her two auxiliary modes, $x$ and
$y$. We denote Eve's decoded information as $k$.

\emph{Eve's optimal strategy} Eve successfully decodes $k=0$ if her
$x$ ancillary spatial mode is in the empty state and her $y$ ancillary
spatial mode is in the state $|0\rangle$, which can be represented
as $|{\rm vac}\rangle_{x}|0\rangle_{y}$. She decodes $k=1$ when
she gets nonempty state in her $x$ ancillary spatial mode or empty
state and $|1\rangle$ in her $x$ and $y$ ancillary spatial modes,
respectively, i.e., $|0\rangle_{x}|{\rm vac}\rangle_{y}$ and $|1\rangle_{x}|{\rm vac}\rangle_{y}$
or $|{\rm vac}\rangle_{x}|1\rangle_{y}$. Now, let's consider a parameter $p_{jmk}^{E_{1}}$,
which characterizes the joint probability where $j$, $m$ and $k$
represent Alice, Bob and Eve's encoding, decoding, and decoding information,
respectively. We can write the joint probabilities using Eqs. (\ref{eq:Final_State_Bob_Zero_Alice_Zero_E1})
$-$ (\ref{eq:Final_State_Bob_Minus_Alice_One}) as follows.

\begin{equation}
\begin{array}{lcl}
p_{000}^{E_{1}} & = & q\\
\\
p_{001}^{E_{1}} & = & p_{010}^{E_{1}}=p_{011}^{E_{1}}=0\\
\\
p_{100}^{E_{1}} & = & \frac{1}{4}\left(1-q\right)\\
\\
p_{101}^{E_{1}} & = & \left(1-q\right)\left(\frac{1}{4}+\frac{p}{2}\right)\\
\\
p_{110}^{E_{1}} & = & 0\\
\\
p_{111}^{E_{1}} & = & \frac{1}{2}\left(1-p\right)\left(1-q\right)
\end{array}.\label{eq:Joint_Probability_E1_Attack}
\end{equation}
We can determine the mutual information among Alice, Bob and Eve
by applying Eq. (\ref{eq:Joint_Probability_E1_Attack}) . To simplify
the resulting expressions, we utilize Shannon entropy represented
as, $\mathbf{H}\left[x\right]=-x\log_{2}x$, where $x$ represents
the probability of an event occurring.

\begin{equation}
\begin{array}{lcl}
H\left({\rm B|A}\right)_{E_{1}} & = & \left(1-q\right)\left(\mathbf{H}\left[\frac{1}{2}\left(1-p\right)\right]+\mathbf{H}\left[\frac{1}{2}\left(1+p\right)\right]\right)\\
\\
H\left({\rm B}\right)_{E_{1}} & = & \mathbf{H}\left[q+\frac{1}{2}\left(1-q\right)\left(1+p\right)\right]+\mathbf{H}\left[\frac{1}{2}\left(1-p\right)\left(1-q\right)\right]\\
\\
I\left({\rm A,B}\right)_{E_{1}} & = & H\left({\rm B}\right)_{E_{1}}-H\left({\rm B|A}\right)_{E_{1}}
\end{array},\label{eq:Mutual_Information_Alice_Bob_E1_Attack}
\end{equation}

\begin{equation}
\begin{array}{lcl}
H\left({\rm A|E}\right)_{E_{1}} & = & \frac{1}{4}\left(1+3q\right)\left(\mathbf{H}\left[\frac{4q}{1+3q}\right]+\mathbf{H}\left[\frac{1-q}{1+3q}\right]\right)\\
\\
H\left({\rm A}\right)_{E_{1}} & = & \mathbf{H}\left[q\right]+\mathbf{H}\left[1-q\right]\\
\\
I\left({\rm A,E}\right)_{E_{1}} & = & H\left({\rm A}\right)_{E_{1}}-H\left({\rm A|E}\right)_{E_{1}}
\end{array},\label{eq:Mutual_Information_Alice_Eve_E1_Attack}
\end{equation}
and

\begin{equation}
\begin{array}{lcl}
H\left({\rm B|E}\right)_{E_{1}} & = & \frac{3}{4}\left(1-q\right)\left(\mathbf{H}\left[\frac{1}{3}\left(1+2p\right)\right]+\mathbf{H}\left[\frac{2}{3}\left(1-p\right)\right]\right)\\
\\
H\left({\rm B}\right)_{E_{1}} & = & \mathbf{H}\left[\frac{1}{2}\left(p+q-pq+1\right)\right]+\mathbf{H}\left[\frac{1}{2}\left(1-p\right)\left(1-q\right)\right]\\
\\
I\left({\rm B,E}\right)_{E_{1}} & = & H\left({\rm B}\right)_{E_{1}}-H\left({\rm B|E}\right)_{E_{1}}
\end{array}.\label{eq:Mutual_Information_Bob_Eve_E1_Attack}
\end{equation}
It is also evident that eavesdropping causes QBER,

\begin{equation}
\begin{array}{lcl}
{\rm QBER}_{E_{1}} & = & \underset{j\ne m}{\sum}p_{jmk}^{E_{1}}\\
 & = & \frac{1}{2}\left(1-q\right)\left(1+p\right)
\end{array}.\label{eq:QBER_E1_Attack}
\end{equation}

\textbf{\emph{Control mode:}} As explained in \cite{DL04}, the ``double''
control mode mentioned earlier consists of two single tests on the
quantum channel, each of which is equivalent to the one performed
in the one-way BB84 protocol \cite{BB84,GRT+02}. After Eve's B-A
attack, when Bob prepares the initial state $|0\rangle_{t}$, the
composite system becomes $|{\rm B-A}\rangle_{|0\rangle E_{1}}=\frac{1}{\sqrt{2}}\left[|0\rangle|0\rangle|{\rm vac}\rangle+|{\rm vac}\rangle|0\rangle|1\rangle\right]_{txy}$.
In the control mode, Alice randomly measures the traveling qubit in
both the computational ($Z$ basis) and diagonal ($X$ basis) bases
and then sends the projected qubit back to Bob. Bob, in turn, measures
the traveling qubit on the same basis he initially prepared (in this
case, the $Z$ basis). Consequently, we will only consider cases where
Alice chooses the $Z$ basis to measure the state of $t$. The expression
$|{\rm B-A}\rangle_{|0\rangle E_{1}}$ indicates that there is a $\frac{1}{2}$
probability of Alice not detecting any photon. However, if the photon
is detected, its state is the same as Bob's initial state\footnote{It implies that there is no detection probability of eavesdropping
in Eve's B-A attack for $Z$ basis scenarios.}. So, the detection probability of eavesdropping based on the measurement
outcome is non-existent. However, eavesdropping can still be identified
by monitoring the losses, as long as the channel's transmittance for
Alice and Bob is below $0.5$. In this scenario, Eve performs an A-B
attack $\left(Q_{txy}^{-1}\right)$ on the photon that Alice detects.
Subsequently, Bob measures the state $t$ in the $Z$ basis and returns
the projected qubit to Alice. The composite system after Eve's A-B
attack on Alice and Bob is as follows,

\begin{equation}
\begin{array}{lcl}
Q_{txy}^{-1}\,|0\rangle_{t}|0\rangle_{x}|{\rm vac}\rangle_{y} & = & H_{y}\,{\rm CPBS}_{txy}\,{\rm SWAP}_{tx}\,|0\rangle_{t}|0\rangle_{x}|{\rm vac}\rangle_{y}\\
 & = & \frac{1}{\sqrt{2}}\left[|0\rangle|{\rm vac}\rangle|0\rangle+|0\rangle|{\rm vac}\rangle|1\rangle\right]_{txy}
\end{array}.\label{eq:Control_Bob_Zero_State_Zero_Zero_Empty}
\end{equation}
When Bob's initial state is $|1\rangle_{t}$, Alice encounters an
equally likely chance of either detecting or not detecting a photon.
Bob measures the state $t$ on $Z$ basis and sends the the projected
qubit back to Alice. The composite system after Eve's A-B attack is,

\begin{equation}
\begin{array}{lcl}
Q_{txy}^{-1}\,|1\rangle_{t}|1\rangle_{x}|{\rm vac}\rangle_{y} & = & \frac{1}{\sqrt{2}}\left[|1\rangle|{\rm vac}\rangle|0\rangle-|1\rangle|{\rm vac}\rangle|1\rangle\right]_{txy}\end{array}.\label{eq:Control_Bob_One_State_One_One_Empty}
\end{equation}

The situation will change considerably if Bob selects the initial
state in the $X$ basis instead of the $Z$ basis. For instance, if
we consider Bob's initial state as $|+\rangle_{t}$, the resulting
composite system after Eve's B-A attack will be,

\[
\begin{array}{lcl}
|{\rm B-A}\rangle_{|+\rangle E_{1}} & = & \frac{1}{2}\left[|0\rangle|0\rangle|{\rm vac}\rangle+|{\rm vac}\rangle|0\rangle|1\rangle+|{\rm vac}\rangle|1\rangle|0\rangle+|1\rangle|1\rangle|{\rm vac}\rangle\right]_{txy}\\
 & = & \frac{1}{2}\left[\frac{1}{\sqrt{2}}\left\{ \left(|+\rangle+|-\rangle\right)|0\rangle|{\rm vac}\rangle+\left(|+\rangle-|-\rangle\right)|1\rangle|{\rm vac}\rangle\right\} +|{\rm vac}\rangle|0\rangle|1\rangle+|{\rm vac}\rangle|1\rangle|0\rangle\right]_{txy}
\end{array}.
\]
Alice also encounters
no detection of a photon with a probability of $\frac{1}{2}$, similar
to the $Z$ basis scenarios. However, in all detected cases, the photon
is not in Bob's initial state ($|+\rangle_{t}$). This implies that
in the B-A attack $\left(Q_{txy}\right)$, Eve's detection occurs
for $X$ basis scenarios, which does not happen in $Z$ basis cases.
Now, we will thoroughly analyze Eve's A-B attack $\left(Q_{txy}^{-1}\right)$
for all four possible outcomes of Alice's measurement outcome. The
final composite system on which Bob performs his measurement, given
the collapsed state after Alice's measurement, is $|+\rangle_{t}|0\rangle_{x}|{\rm vac}\rangle_{y}$,

\begin{equation}
\begin{array}{lcl}
Q_{txy}^{-1}\,|+\rangle_{t}|0\rangle_{x}|{\rm vac}\rangle_{y} & = & Q_{txy}^{-1}\,\frac{1}{\sqrt{2}}\left[|0\rangle|0\rangle|{\rm vac}\rangle+|1\rangle|0\rangle|{\rm vac}\rangle\right]_{txy}\\
 & = & \frac{1}{\sqrt{2}}|0\rangle_{t}\left[\frac{1}{\sqrt{2}}|{\rm vac}\rangle\left(|0\rangle+|1\rangle\right)+|1\rangle|{\rm vac}\rangle\right]_{xy}\\
 & = & \frac{1}{2}\left(|+\rangle+|-\rangle\right)_{t}\left[\frac{1}{\sqrt{2}}|{\rm vac}\rangle\left(|0\rangle+|1\rangle\right)+|1\rangle|{\rm vac}\rangle\right]_{xy}
\end{array}.\label{eq:Control_Bob_Plus_State_Plus_Zero_Empty}
\end{equation}
Similarly, the rest of three composite systems after Alice's measurement
are $|+\rangle_{t}|1\rangle_{x}|{\rm vac}\rangle_{y}$, $|-\rangle_{t}|0\rangle_{x}|{\rm vac}\rangle_{y}$,
and $|-\rangle_{t}|1\rangle_{x}|{\rm vac}\rangle_{y}$. Consequently,
the final composite systems are,

\begin{equation}
\begin{array}{lcl}
Q_{txy}^{-1}\,|+\rangle_{t}|1\rangle_{x}|{\rm vac}\rangle_{y} & = & \frac{1}{2}\left(|+\rangle-|-\rangle\right)_{t}\left[\frac{1}{\sqrt{2}}|{\rm vac}\rangle\left(|0\rangle-|1\rangle\right)+|0\rangle|{\rm vac}\rangle\right]_{xy}\end{array},\label{eq:Control_Bob_Plus_State_Plus_One_Empty}
\end{equation}

\begin{equation}
\begin{array}{lcl}
Q_{txy}^{-1}\,|-\rangle_{t}|0\rangle_{x}|{\rm vac}\rangle_{y} & = & \frac{1}{2}\left(|+\rangle+|-\rangle\right)_{t}\left[\frac{1}{\sqrt{2}}|{\rm vac}\rangle\left(|0\rangle+|1\rangle\right)-|1\rangle|{\rm vac}\rangle\right]_{xy}\end{array},\label{eq:Control_Bob_Plus_State_Minus_Zero_Empty}
\end{equation}
and

\begin{equation}
\begin{array}{lcl}
Q_{txy}^{-1}\,|-\rangle_{t}|1\rangle_{x}|{\rm vac}\rangle_{y} & = & \frac{1}{2}\left(|+\rangle-|-\rangle\right)_{t}\left[-\frac{1}{\sqrt{2}}|{\rm vac}\rangle\left(|0\rangle-|1\rangle\right)+|0\rangle|{\rm vac}\rangle\right]_{xy}\end{array}.\label{eq:Control_Bob_Plus_State_Minus_One_Empty}
\end{equation}
If Bob's initial state is $|-\rangle_{t}$, the composite state after
Eve's B-A attack is,

\[
\begin{array}{lcl}
|{\rm B-A}\rangle_{|-\rangle E_{1}} & = & \frac{1}{2}\left[|0\rangle|0\rangle|{\rm vac}\rangle+|{\rm vac}\rangle|0\rangle|1\rangle-|{\rm vac}\rangle|1\rangle|0\rangle-|1\rangle|1\rangle|{\rm vac}\rangle\right]_{txy}\\
 & = & \frac{1}{2}\left[\frac{1}{\sqrt{2}}\left\{ \left(|+\rangle+|-\rangle\right)|0\rangle|{\rm vac}\rangle-\left(|+\rangle-|-\rangle\right)|1\rangle|{\rm vac}\rangle\right\} +|{\rm vac}\rangle|0\rangle|1\rangle-|{\rm vac}\rangle|1\rangle|0\rangle\right]_{txy}
\end{array}
\]
From an intuitive perspective, it becomes evident that the entire
scenario appears analogous whether Bob's initial state is $|-\rangle_{t}$
or $|+\rangle_{t}$. Therefore, Eve's A-B attack will have the same
impact on all four different outcomes of Alice's measurement, leading
to identical results as described by the Eqs. (\ref{eq:Control_Bob_Plus_State_Plus_Zero_Empty})
$-$ (\ref{eq:Control_Bob_Plus_State_Minus_One_Empty}).

If we ignore the occurrence of no detection of photons for the optimal
choice of eavesdropping (in favor of Eve), we get the probability
of undetected Eve in $E_{1}$ attack scenario using Eqs. (\ref{eq:Control_Bob_Zero_State_Zero_Zero_Empty})
$-$ (\ref{eq:Control_Bob_Plus_State_Minus_One_Empty}),

\[
\begin{array}{lcl}
P_{nd}^{E_{1}} & = & \frac{1}{4}\left(1+1+\frac{1}{4}+\frac{1}{4}\right)=\frac{5}{8}\end{array},
\]
the average\footnote{Eve has two attack methods: B-A attack and A-B attack. And there are
``double'' control mode for Eve's detection \cite{DL04} for which
$\frac{1}{2}$ factor arises.} detection probability of Eve's presence in $E_{1}$ attack scenario
is,

\begin{equation}
\begin{array}{lcl}
P_{d}^{E_{1}} & = & \frac{1}{2}\left(1-\frac{5}{8}\right)=\frac{3}{16}=0.1875\end{array}.\label{eq:Eve_Detection_Probability_E1_Attack}
\end{equation}
Alice and Bob can tolerate the maximum detection probability of Eve's
presence in control mode for $E_{1}$ attack scenario is $0.1875$.
In other words, the threshold bound for the probability of detecting
Eve is $18.75\%$.

\section*{Appendix B}

\textbf{\emph{Message mode:}} W{\'o}jcik\textcolor{red}{{} }introduced
an alternative method for a symmetry attack (referred to as $E_{2}$),
where, with $\frac{1}{2}$ probability, an additional unitary operation
$S_{ty}$ is applied right after the operation $Q_{txy}^{-1}$ during
the A-B attack \cite{W03}. The $S_{ty}$ operation is defined as
$X_{t}Z_{t}{\rm CNOT}_{ty}X_{t}$, combining $Z$, negation $X$,
and controlled negation (CNOT) operations. The B-A attack remains
the same as in $E_{1}$, with the modification exclusively applied
to the B-A attack. The notation (superscripts and subscripts) is consistent
with the previous notation, with the only change being $E_{1}$ replaced
by $E_{2}$. The composite system's final state, after the $S_{ty}$
operation is performed in the A-B attack when Alice encodes a $0$
, and Bob's initial state is $|0\rangle_{t}$ would be

\begin{equation}
\begin{array}{lcl}
|{\rm A-B}\rangle_{|0\rangle E_{2}}^{0} & = & S_{ty}\,|{\rm A-B}\rangle_{|0\rangle E_{1}}^{0}\\
 & = & X_{t}\,Z_{t}\,{\rm CNOT}_{ty}\,X_{t}\,|0\rangle_{t}|{\rm vac}\rangle_{x}|0\rangle_{y}\\
 & = & X_{t}\,Z_{t}\,{\rm CNOT}_{ty}\,|1\rangle_{t}|{\rm vac}\rangle_{x}|0\rangle_{y}\\
 & = & X_{t}\,Z_{t}\,|1\rangle_{t}|{\rm vac}\rangle_{x}|0\rangle_{y}\\
 & = & X_{t}\,\left(-|1\rangle_{t}|{\rm vac}\rangle_{x}|0\rangle_{y}\right)\\
 & =- & |0\rangle_{t}|{\rm vac}\rangle_{x}|0\rangle_{y}
\end{array}.\label{eq:Final_State_Bob_Zero_Alice_Zero_E2}
\end{equation}
Likewise, we can get the remaining composite systems in which Alice
encodes $0$ when Bob's initial states are $|1\rangle$, $|+\rangle$,
and $|-\rangle$, i.e.,

\begin{equation}
\begin{array}{lcl}
|{\rm A-B}\rangle_{|1\rangle E_{2}}^{0} & = & |1\rangle_{t}|{\rm vac}\rangle_{x}|0\rangle_{y}\end{array},\label{eq:Final_State_Bob_One_Alice_Zero_E2}
\end{equation}

\begin{equation}
\begin{array}{lcl}
|{\rm A-B}\rangle_{|+\rangle E_{2}}^{0} & = & \frac{1}{2}\left[-|+\rangle|{\rm vac}\rangle|1\rangle-|-\rangle|{\rm vac}\rangle|1\rangle+|+\rangle|{\rm vac}\rangle|0\rangle-|-\rangle|{\rm vac}\rangle|0\rangle\right]_{txy}\end{array},\label{eq:Final_State_Bob_Plus_Alice_Zero_E2}
\end{equation}
and

\begin{equation}
\begin{array}{lcl}
|{\rm A-B}\rangle_{|-\rangle E_{2}}^{0} & = & \frac{1}{2}\left[-|+\rangle|{\rm vac}\rangle|1\rangle-|-\rangle|{\rm vac}\rangle|1\rangle-|+\rangle|{\rm vac}\rangle|0\rangle+|-\rangle|{\rm vac}\rangle|0\rangle\right]_{txy}\end{array},\label{eq:Final_State_Bob_Minus_Alice_Zero_E2}
\end{equation}
respectively.

The final composite system after performing $S_{ty}$ operation in
A-B attack when Alice encodes $1$ and Bob's initial state is $|0\rangle_{t}$
would be

\begin{equation}
\begin{array}{lcl}
|{\rm A-B}\rangle_{|0\rangle E_{2}}^{1} & = & S_{ty}\,|{\rm A-B}\rangle_{|0\rangle E_{1}}^{1}\\
 & = & X_{t}\,Z_{t}\,{\rm CNOT}_{ty}\,X_{t}\,\left[-\frac{1}{\sqrt{2}}|0\rangle|1\rangle|{\rm vac}\rangle+\frac{1}{2}|0\rangle|{\rm vac}\rangle|0\rangle-\frac{1}{2}|0\rangle|{\rm vac}\rangle|1\rangle\right]_{txy}\\
 & = & X_{t}\,Z_{t}\,{\rm CNOT}_{ty}\,\left[-\frac{1}{\sqrt{2}}|1\rangle|1\rangle|{\rm vac}\rangle+\frac{1}{2}|1\rangle|{\rm vac}\rangle|0\rangle-\frac{1}{2}|1\rangle|{\rm vac}\rangle|1\rangle\right]_{txy}\\
 & = & X_{t}\,Z_{t}\,\left[-\frac{1}{\sqrt{2}}|1\rangle|1\rangle|{\rm vac}\rangle+\frac{1}{2}|1\rangle|{\rm vac}\rangle|1\rangle-\frac{1}{2}|1\rangle|{\rm vac}\rangle|0\rangle\right]_{txy}\\
 & = & X_{t}\,\left[\frac{1}{\sqrt{2}}|1\rangle|1\rangle|{\rm vac}\rangle-\frac{1}{2}|1\rangle|{\rm vac}\rangle|1\rangle+\frac{1}{2}|1\rangle|{\rm vac}\rangle|0\rangle\right]_{txy}\\
 & = & \left[\frac{1}{\sqrt{2}}|0\rangle|1\rangle|{\rm vac}\rangle-\frac{1}{2}|0\rangle|{\rm vac}\rangle|1\rangle+\frac{1}{2}|0\rangle|{\rm vac}\rangle|0\rangle\right]_{txy}
\end{array}.\label{eq:Final_State_Bob_Zero_Alice_One_E2}
\end{equation}
Similarly, we can describe the remaining composite systems in which
Alice encodes $1$ when Bob's initial states are $|1\rangle$, $|+\rangle$
and $|-\rangle$, i.e.,

\begin{equation}
\begin{array}{lcl}
|{\rm A-B}\rangle_{|1\rangle E_{2}}^{1} & = & \left[\frac{1}{\sqrt{2}}|1\rangle|0\rangle|{\rm vac}\rangle+\frac{1}{2}|1\rangle|{\rm vac}\rangle|0\rangle+\frac{1}{2}|1\rangle|{\rm vac}\rangle|1\rangle\right]\end{array},\label{eq:Final_State_Bob_One_Alice_One_E2}
\end{equation}

\begin{equation}
\begin{array}{lcl}
|{\rm A-B}\rangle_{|+\rangle E_{2}}^{1} & = & \frac{1}{2}\left[-|-\rangle|{\rm vac}\rangle|1\rangle+|+\rangle|{\rm vac}\rangle|0\rangle\right]_{txy}+\frac{1}{2\sqrt{2}}\left[|-\rangle|1\rangle|{\rm vac}\rangle+|+\rangle|1\rangle|{\rm vac}\rangle-|-\rangle|0\rangle|{\rm vac}\rangle+|+\rangle|0\rangle|{\rm vac}\rangle\right]_{txy}\end{array},\label{eq:Final_State_Bob_Plus_Alice_One_E2}
\end{equation}
and

\begin{equation}
\begin{array}{lcl}
|{\rm A-B}\rangle_{|-\rangle E_{2}}^{1} & = & \frac{1}{2}\left[-|+\rangle|{\rm vac}\rangle|1\rangle+|-\rangle|{\rm vac}\rangle|0\rangle\right]_{txy}+\frac{1}{2\sqrt{2}}\left[|-\rangle|1\rangle|{\rm vac}\rangle+|+\rangle|1\rangle|{\rm vac}\rangle+|-\rangle|0\rangle|{\rm vac}\rangle-|+\rangle|0\rangle|{\rm vac}\rangle\right]_{txy}\end{array},\label{eq:Final_State_Bob_Minus_Alice_One_E2}
\end{equation}
respectively.

\emph{Eve's optimal strategy} The optimal strategy for Eve to decipher
Alice's encoded information, following her measurement of the $x$
and $y$ ancillary spatial modes, aligns with the $E_{1}$ attack
scenario. We have already mentioned that with $\frac{1}{2}$ probability
of the additional unitary operation $S_{ty}$ is performed. Consequently,
Eq. (\ref{eq:Joint_Probability_E1_Attack}) remains applicable for
computing joint probabilities in half of all conceivable scenarios.
These joint probabilities, denoted as $p_{jmk}^{E_{2}}$, where $j$,
$m$ and $k$ represent Alice, Bob and Eve's encoding, decoding
and decoding information, respectively. We can write the joint probabilities
using Eqs. (\ref{eq:Joint_Probability_E1_Attack}) and (\ref{eq:Final_State_Bob_Zero_Alice_Zero_E2})
$-$ (\ref{eq:Final_State_Bob_Minus_Alice_One_E2}),

\begin{equation}
\begin{array}{lcl}
p_{000}^{E_{2}} & = & \frac{1}{2}\left[\frac{q}{4}\left(1+p\right)+q\right]=\frac{q}{8}\left(5+p\right)\\
\\
p_{001}^{E_{2}} & = & \frac{q}{8}\left(1+p\right)\\
\\
p_{010}^{E_{2}} & = & p_{011}^{E_{2}}=\frac{q}{8}\left(1-p\right)\\
\\
p_{100}^{E_{2}} & = & \frac{1}{4}\left(1-q\right)\\
\\
p_{101}^{E_{2}} & =\frac{1}{4} & \left(1-q\right)\left(1+2p\right)\\
\\
p_{110}^{E_{2}} & = & 0\\
\\
p_{111}^{E_{2}} & = & \frac{1}{2}\left(1-p\right)\left(1-q\right)
\end{array}.\label{eq:Joint_Probability_E2_Attack}
\end{equation}
By utilizing Equation (\ref{eq:Joint_Probability_E2_Attack}), we
can derive the mutual information expression among Alice, Bob and
Eve. This can be represented as follows,

\begin{equation}
\begin{array}{lcl}
H\left({\rm B|A}\right)_{E_{2}} & = & q\left(\mathbf{H}\left[\frac{1}{4}\left(3+p\right)\right]+\mathbf{H}\left[\frac{1}{4}\left(1-p\right)\right]\right)+\left(1-q\right)\left(\mathbf{H}\left[\frac{1}{2}\left(1+p\right)\right]+\mathbf{H}\left[\frac{1}{2}\left(1-p\right)\right]\right)\\
\\
H\left({\rm B}\right)_{E_{2}} & = & \mathbf{H}\left[\frac{1}{4}\left(2+2p+q-pq+2\right)\right]+\mathbf{H}\left[\frac{1}{4}\left(1-p\right)\left(2-q\right)\right]\\
\\
I\left({\rm A,B}\right)_{E_{2}} & = & H\left({\rm B}\right)_{E_{2}}-H\left({\rm B|A}\right)_{E_{2}}
\end{array},\label{eq:Mutual_Information_Alice_Bob_E2_Attack}
\end{equation}

\begin{equation}
\begin{array}{lcl}
H\left({\rm E|A}\right)_{E_{2}} & = & 0.811278\\
\\
H\left({\rm E}\right)_{E_{2}} & = & \mathbf{H}\left[\frac{1}{4}\left(1+2q\right)\right]+\mathbf{H}\left[\frac{1}{4}\left(3-2q\right)\right]\\
\\
I\left({\rm A,E}\right)_{E_{2}} & = & H\left({\rm E}\right)_{E_{2}}-H\left({\rm E|A}\right)_{E_{2}}
\end{array},\label{eq:Mutual_Information_Alice_Eve_E2_Attack}
\end{equation}
and

\begin{equation}
\begin{array}{lcl}
H\left({\rm E|B}\right)_{E_{2}} & = & \frac{1}{4}\left(2+2p+q-pq\right)\left(\mathbf{H}\left[\frac{1}{2}\left(\frac{2+3q+pq}{2+2p+q-pq}\right)\right]+\mathbf{H}\left[\frac{1}{2}\left(\frac{2+4p-q-3pq}{2+2p+q-pq}\right)\right]\right)\\
 & + & \frac{1}{4}\left(1-p\right)\left(2-q\right)\left(\mathbf{H}\left[\frac{1}{2}\left(\frac{q}{2-q}\right)\right]+\mathbf{H}\left[\frac{1}{2}\left(\frac{4-3q}{2-q}\right)\right]\right)\\
\\
H\left({\rm E}\right)_{E_{2}} & = & \mathbf{H}\left[\frac{1}{4}\left(1+2q\right)\right]+\mathbf{H}\left[\frac{1}{4}\left(3-2q\right)\right]\\
\\
I\left({\rm B,E}\right)_{E_{2}} & = & H\left({\rm E}\right)_{E_{2}}-H\left({\rm E|B}\right)_{E_{2}}
\end{array}.\label{eq:Mutual_Information_Bob_Eve_E2_Attack}
\end{equation}
It is also evident that eavesdropping causes QBER,

\begin{equation}
\begin{array}{lcl}
{\rm QBER}_{E_{2}} & = & \underset{j\ne m}{\sum}p_{jmk}^{E_{2}}\\
 & = & \frac{1}{4}\left(2+2p-q-3pq\right)
\end{array}.\label{eq:QBER_E2_Attack}
\end{equation}

\textbf{\emph{Control mode:}} The basic idea of control mode is already
illustrated and explained in $E_{1}$ attack scenario. The primary
distinction lies in Eve's action of applying the $S_{ty}$ operator
with $\frac{1}{2}$ probability following the operation of $Q_{txy}^{-1}$
in the A-B attack. We take the event where Bob prepares $|0\rangle_{t}$
state and Alice measures in the $Z$ basis. So, the composite system
$|B-A\rangle_{|0\rangle E_{1}}$ collapses into the state $|0\rangle_{t}|0\rangle_{x}|{\rm vac}\rangle_{y}$
when photon is detected by Alice with $\frac{1}{2}$ probability.
After Eve's $E_{2}$ attack, the composite system will be,

\begin{equation}
\begin{array}{lcl}
S_{ty}Q_{txy}^{-1}\,|0\rangle_{t}|0\rangle_{x}|{\rm vac}\rangle_{y} & = & S_{ty}\,H_{y}\,{\rm CPBS}_{txy}\,{\rm SWAP}_{tx}\,|0\rangle_{t}|0\rangle_{x}|{\rm vac}\rangle_{y}\\
 & = & S_{ty}\,\frac{1}{\sqrt{2}}\left[|0\rangle|{\rm vac}\rangle|0\rangle+|0\rangle|{\rm vac}\rangle|1\rangle\right]_{txy}\\
 & = & X_{t}\,Z_{t}\,{\rm CNOT}_{ty}\,X_{t}\,\frac{1}{\sqrt{2}}\left[|0\rangle|{\rm vac}\rangle|0\rangle+|0\rangle|{\rm vac}\rangle|1\rangle\right]_{txy}\\
 & = & X_{t}\,Z_{t}\,{\rm CNOT}_{ty}\,\frac{1}{\sqrt{2}}\left[|1\rangle|{\rm vac}\rangle|0\rangle+|1\rangle|{\rm vac}\rangle|1\rangle\right]_{txy}\\
 & = & X_{t}\,Z_{t}\,\frac{1}{\sqrt{2}}\left[|1\rangle|{\rm vac}\rangle|1\rangle+|1\rangle|{\rm vac}\rangle|0\rangle\right]_{txy}\\
 & = & X_{t}\,\frac{1}{\sqrt{2}}\left[-|1\rangle|{\rm vac}\rangle|1\rangle-|1\rangle|{\rm vac}\rangle|0\rangle\right]_{txy}\\
 & = & \frac{1}{\sqrt{2}}\left[-|0\rangle|{\rm vac}\rangle|1\rangle-|0\rangle|{\rm vac}\rangle|0\rangle\right]_{txy}
\end{array}.\label{eq:Control_E2_Bob_Zero_State_Zero_Zero_Empty}
\end{equation}
When Bob's initial state is $|1\rangle_{t}$, then Alice also encounters
an equal probability of photon detection and non-detection. Alice
proceeds to measure the state $t$ on $Z$ basis and subsequently
sends the the projected qubit back to Bob. The composite system after
Eve's A-B attack is,

\begin{equation}
\begin{array}{lcl}
S_{ty}Q_{txy}^{-1}\,|1\rangle_{t}|1\rangle_{x}|{\rm vac}\rangle_{y} & = & \frac{1}{\sqrt{2}}\left[|1\rangle|{\rm vac}\rangle|0\rangle-|1\rangle|{\rm vac}\rangle|1\rangle\right]_{txy}\end{array}.\label{eq:Control_E2_Bob_One_State_One_One_Empty}
\end{equation}

Now, let's consider Bob's initial state, denoted as, $|+\rangle_{t}$,
the composite system after Eve's B-A attack is identical to $|{\rm B-A}\rangle_{|+\rangle E_{1}}$.
When Alice performs collapse measurement operation by $X$ basis measurement,
there will be a total of four composite states, i.e., $|+\rangle_{t}|0\rangle_{x}|{\rm vac}\rangle_{y}$,
$|-\rangle_{t}|0\rangle_{x}|{\rm vac}\rangle_{y}$, $|+\rangle_{t}|1\rangle_{x}|{\rm vac}\rangle_{y}$,
and $|-\rangle_{t}|1\rangle_{x}|{\rm vac}\rangle_{y}$. We can write
from Eqs. (\ref{eq:Control_Bob_Plus_State_Plus_Zero_Empty}) $-$
(\ref{eq:Control_Bob_Plus_State_Minus_One_Empty}),

\begin{equation}
\begin{array}{lcl}
S_{ty}\,Q_{txy}^{-1}\,|+\rangle_{t}|0\rangle_{x}|{\rm vac}\rangle_{y} & = & S_{ty}\,\frac{1}{2}\left(|+\rangle+|-\rangle\right)_{t}\left[\frac{1}{\sqrt{2}}|{\rm vac}\rangle\left(|0\rangle+|1\rangle\right)+|1\rangle|{\rm vac}\rangle\right]_{xy}\\
 & = & X_{t}\,Z_{t}\,{\rm CNOT}_{ty}\,X_{t}\,\left[\frac{1}{2}\left(|0\rangle|{\rm vac}\rangle|0\rangle+|0\rangle|{\rm vac}\rangle|1\rangle\right)+\frac{1}{\sqrt{2}}|0\rangle|1\rangle|{\rm vac}\rangle\right]_{txy}\\
 & = & X_{t}\,Z_{t}\,\left[\frac{1}{2}\left(|1\rangle|{\rm vac}\rangle|1\rangle+|1\rangle|{\rm vac}\rangle|0\rangle\right)+\frac{1}{\sqrt{2}}|1\rangle|1\rangle|{\rm vac}\rangle\right]_{txy}\\
 & = & \left[\frac{1}{2}\left(-|0\rangle|{\rm vac}\rangle|1\rangle-|0\rangle|{\rm vac}\rangle|0\rangle\right)-\frac{1}{\sqrt{2}}|0\rangle|1\rangle|{\rm vac}\rangle\right]_{txy}\\
 & = & \left[\frac{1}{2\sqrt{2}}\left(-|+\rangle|{\rm vac}\rangle|1\rangle-|-\rangle|{\rm vac}\rangle|1\rangle-|+\rangle|{\rm vac}\rangle|0\rangle-|-\rangle|{\rm vac}\rangle|0\rangle\right)\right.\\
 & - & \left.\frac{1}{2}\left(|+\rangle|1\rangle|{\rm vac}\rangle+|-\rangle|1\rangle|{\rm vac}\rangle\right)\right]_{txy}
\end{array},\label{eq:Control_E2_Bob_Plus_State_Plus_Zero_Empty}
\end{equation}

\begin{equation}
\begin{array}{lcl}
S_{ty}\,Q_{txy}^{-1}\,|+\rangle_{t}|1\rangle_{x}|{\rm vac}\rangle_{y} & = & S_{ty}\,\frac{1}{2}\left(|+\rangle-|-\rangle\right)_{t}\left[\frac{1}{\sqrt{2}}|{\rm vac}\rangle\left(|0\rangle-|1\rangle\right)+|0\rangle|{\rm vac}\rangle\right]_{xy}\\
 & = & X_{t}\,Z_{t}\,{\rm CNOT}_{ty}\,X_{t}\,\left[\frac{1}{2}\left(|1\rangle|{\rm vac}\rangle|0\rangle+|1\rangle|{\rm vac}\rangle|1\rangle\right)+\frac{1}{\sqrt{2}}|1\rangle|0\rangle|{\rm vac}\rangle\right]_{txy}\\
 & = & X_{t}\,Z_{t}\,\left[\frac{1}{2}\left(|0\rangle|{\rm vac}\rangle|0\rangle+|0\rangle|{\rm vac}\rangle|1\rangle\right)+\frac{1}{\sqrt{2}}|0\rangle|0\rangle|{\rm vac}\rangle\right]_{txy}\\
 & = & \left[\frac{1}{2}\left(|1\rangle|{\rm vac}\rangle|0\rangle+|1\rangle|{\rm vac}\rangle|1\rangle\right)+\frac{1}{\sqrt{2}}|1\rangle|0\rangle|{\rm vac}\rangle\right]_{txy}\\
 & = & \left[\frac{1}{2\sqrt{2}}\left(|+\rangle|{\rm vac}\rangle|0\rangle-|-\rangle|{\rm vac}\rangle|0\rangle+|+\rangle|{\rm vac}\rangle|1\rangle-|-\rangle|{\rm vac}\rangle|1\rangle\right)\right.\\
 & + & \left.\frac{1}{2}\left(|+\rangle|0\rangle|{\rm vac}\rangle-|-\rangle|0\rangle|{\rm vac}\rangle\right)\right]_{txy}
\end{array},\label{eq:Control_E2_Bob_Plus_State_Plus_One_Empty}
\end{equation}

\begin{equation}
\begin{array}{lcl}
S_{ty}\,Q_{txy}^{-1}\,|-\rangle_{t}|0\rangle_{x}|{\rm vac}\rangle_{y} & = & S_{ty}\,\frac{1}{2}\left(|+\rangle+|-\rangle\right)_{t}\left[\frac{1}{\sqrt{2}}|{\rm vac}\rangle\left(|0\rangle+|1\rangle\right)-|1\rangle|{\rm vac}\rangle\right]_{xy}\\
 & = & \left[\frac{1}{2\sqrt{2}}\left(-|+\rangle|{\rm vac}\rangle|1\rangle-|-\rangle|{\rm vac}\rangle|1\rangle-|+\rangle|{\rm vac}\rangle|0\rangle-|-\rangle|{\rm vac}\rangle|0\rangle\right)\right.\\
 & + & \left.\frac{1}{2}\left(|+\rangle|1\rangle|{\rm vac}\rangle+|-\rangle|1\rangle|{\rm vac}\rangle\right)\right]_{txy}
\end{array},\label{eq:Control_E2_Bob_Plus_State_Minus_Zero_Empty}
\end{equation}
and

\begin{equation}
\begin{array}{lcl}
S_{ty}\,Q_{txy}^{-1}\,|-\rangle_{t}|1\rangle_{x}|{\rm vac}\rangle_{y} & = & S_{ty}\,\frac{1}{2}\left(|+\rangle-|-\rangle\right)_{t}\left[-\frac{1}{\sqrt{2}}|{\rm vac}\rangle\left(|0\rangle-|1\rangle\right)+|0\rangle|{\rm vac}\rangle\right]_{xy}\\
 & = & \left[-\frac{1}{2\sqrt{2}}\left(|+\rangle|{\rm vac}\rangle|0\rangle-|-\rangle|{\rm vac}\rangle|0\rangle-|+\rangle|{\rm vac}\rangle|1\rangle+|-\rangle|{\rm vac}\rangle|1\rangle\right)\right.\\
 & + & \left.\frac{1}{2}\left(|+\rangle|0\rangle|{\rm vac}\rangle-|-\rangle|0\rangle|{\rm vac}\rangle\right)\right]_{txy}
\end{array},\label{eq:Control_E2_Bob_Plus_State_Minus_One_Empty}
\end{equation}
respectively.

Now, let's consider Bob's initial state $|-\rangle_{t}$. The composite
system after Eve's B-A attack remains unchanged with the state $|{\rm B-A}\rangle_{|-\rangle E_{1}}$.
From an intuitive perspective, it becomes evident that the entire
scenario appears analogous whether Bob's initial state is $|-\rangle_{t}$
or $|+\rangle_{t}$. Therefore, Eve's A-B attack will have the same
impact on all four different outcomes of Alice's measurement, leading
to identical results as described by Eqs. (\ref{eq:Control_E2_Bob_Plus_State_Plus_Zero_Empty})
$-$ (\ref{eq:Control_E2_Bob_Plus_State_Minus_One_Empty}).

If we ignore the occurrence of no detection of photons for the optimal
choice of eavesdropping (in favor of Eve), we get the probability
of undetected Eve in the $E_{2}$ attack scenario using Eqs. (\ref{eq:Control_E2_Bob_Zero_State_Zero_Zero_Empty})
$-$ (\ref{eq:Control_E2_Bob_Plus_State_Minus_One_Empty}),

\[
\begin{array}{lcl}
P_{nd}^{E_{2}} & = & \frac{1}{4}\left(1+1+\frac{1}{4}+\frac{1}{4}\right)=\frac{5}{8}\end{array},
\]
the average detection probability of Eve's presence in $E_{2}$ attack
scenario is,

\begin{equation}
\begin{array}{lcl}
P_{d}^{E_{2}} & = & \frac{1}{2}\left(1-\frac{5}{8}\right)=\frac{3}{16}=0.1875\end{array}.\label{eq:Eve_Detection_Probability_E2_Attack}
\end{equation}
Alice and Bob can tolerate the maximum detection probability of Eve's
presence in control mode for $E_{2}$ attack scenario is $0.1875$.
In other words, the threshold bound for the probability of detecting
Eve is $18.75\%$.

\section*{Appendix C}

Here, we use another attack strategy, $E_{3}$ which was proposed
by Pavi{\v{c}}i{\'c} \cite{P+13}. Eve prepares two auxiliary modes,
denoted as $x$ and $y$, each initially having a single ancilla photon
in the state $|{\rm vac}\rangle_{x}|0\rangle_{y}$, where $|{\rm vac}\rangle$
represents an empty mode. She then directs the traveling photon to
interact with these prepared modes\footnote{In Ref. \cite{P+13}, the author considers qutrit but we use the for
qubit scenario only. For simplicity we use $|0\rangle$ and $|1\rangle$
states rather $|H\rangle$ and $|V\rangle$.}. Similar to the prior two attacks, Eve executes the unitary operation
$Q_{txy}^{\prime}$ when the photon is moving from Bob to Alice in
the B-A attack scenario. Conversely, in the A-B attack situation,
Eve applies the inverse of $Q_{txy}^{\prime}$ $\left(Q_{txy}^{\prime-1}\right)$
to the traveling photon ($t$).

\[
\begin{array}{lcl}
Q_{txy}^{\prime} & = & {\rm CNOT}_{ty}\left({\rm CNOT}_{tx}\otimes I_{y}\right)\left(I_{t}\otimes{\rm PBS}_{xy}\right){\rm CNOT}_{ty}\left({\rm CNOT}_{tx}\otimes I_{y}\right)\left(I_{t}\otimes H_{x}\otimes H_{y}\right)\end{array}.
\]
Eve first applies Hadamard gates to the $x$ and $y$ mode, then two
CNOT gates, a polarizing beam splitter (PBS) gate and two more CNOT
gates (see Ref. \cite{P+13} for more details).

\textbf{\emph{Message mode:}} Let's consider the scenarios where Alice
encodes $1$. In particular, we consider the situation where Bob prepares
the initial state $|0\rangle_{t}$. We will then describe Eve's attack,
known as the B-A attack, on the composite system $|0\rangle_{t}|{\rm vac}\rangle_{x}|0\rangle_{y}$,
which is associated with the $E_{3}$ attack strategy,

\[
\begin{array}{lcl}
|{\rm B-A}\rangle_{|0\rangle E_{3}} & = & Q_{txy}^{\prime}\,|0\rangle_{t}|{\rm vac}\rangle_{x}|0\rangle_{y}\\
 & = & {\rm CNOT}_{ty}\left({\rm CNOT}_{tx}\otimes I_{y}\right)\left(I_{t}\otimes{\rm PBS}_{xy}\right){\rm CNOT}_{ty}\left({\rm CNOT}_{tx}\otimes I_{y}\right)\left(I_{t}\otimes H_{x}\otimes H_{y}\right)\,\left(|0\rangle|{\rm vac}\rangle|0\rangle\right)_{txy}\\
 & = & {\rm CNOT}_{ty}\left({\rm CNOT}_{tx}\otimes I_{y}\right)\left(I_{t}\otimes{\rm PBS}_{xy}\right){\rm CNOT}_{ty}\left({\rm CNOT}_{tx}\otimes I_{y}\right)\,\frac{1}{\sqrt{2}}\left[|0\rangle|{\rm vac}\rangle\left(|0\rangle+|1\rangle\right)\right]_{txy}\\
 & = & {\rm CNOT}_{ty}\left({\rm CNOT}_{tx}\otimes I_{y}\right)\left(I_{t}\otimes{\rm PBS}_{xy}\right){\rm CNOT}_{ty}\,\frac{1}{\sqrt{2}}\left[|0\rangle|{\rm vac}\rangle\left(|0\rangle+|1\rangle\right)\right]_{txy}\\
 & = & {\rm CNOT}_{ty}\left({\rm CNOT}_{tx}\otimes I_{y}\right)\left(I_{t}\otimes{\rm PBS}_{xy}\right)\,\frac{1}{\sqrt{2}}\left[|0\rangle|{\rm vac}\rangle\left(|0\rangle+|1\rangle\right)\right]_{txy}\\
 & = & {\rm CNOT}_{ty}\left({\rm CNOT}_{tx}\otimes I_{y}\right)\,\frac{1}{\sqrt{2}}\left[|0\rangle|0\rangle|{\rm vac}\rangle+|0\rangle|{\rm vac}\rangle|1\rangle\right]_{txy}\\
 & = & {\rm CNOT}_{ty}\,\frac{1}{\sqrt{2}}\left[|0\rangle|0\rangle|{\rm vac}\rangle+|0\rangle|{\rm vac}\rangle|1\rangle\right]_{txy}\\
 & = & \frac{1}{\sqrt{2}}\left[|0\rangle|0\rangle|{\rm vac}\rangle+|0\rangle|{\rm vac}\rangle|1\rangle\right]_{txy}
\end{array},
\]
as Alice performs the operation $iY_{t}^{1}$ (to encode $1$) on
the state $t$, the composite system $|{\rm B-A}\rangle_{|0\rangle E_{3}}$will
undergo a transformation, resulting in a new system denoted as $|{\rm B-A}\rangle_{|0\rangle E_{3}}^{1}$,
where the superscript $1$ signifies that this system now contains
the bit information encoded by Alice. Following Eve's attack $\left(Q_{txy}^{\prime-1}\right)$,
the system has been altered as,

\begin{equation}
\begin{array}{lcl}
|{\rm A-B}\rangle_{|0\rangle E_{3}}^{1} & = & Q_{txy}^{\prime-1}\,\frac{-1}{\sqrt{2}}\left[|1\rangle|0\rangle|{\rm vac}\rangle+|1\rangle|{\rm vac}\rangle|1\rangle\right]_{txy}\\
 & = & \left(I_{t}\otimes H_{x}\otimes H_{y}\right)\left({\rm CNOT}_{tx}\otimes I_{y}\right){\rm CNOT}_{ty}\left(I_{t}\otimes{\rm PBS}_{xy}\right)\\
 & \times & \left({\rm CNOT}_{tx}\otimes I_{y}\right){\rm CNOT}_{ty}\,\frac{-1}{\sqrt{2}}\left[|1\rangle|0\rangle|{\rm vac}\rangle+|1\rangle|{\rm vac}\rangle|1\rangle\right]_{txy}\\
 & = & \left(I_{t}\otimes H_{x}\otimes H_{y}\right)\left({\rm CNOT}_{tx}\otimes I_{y}\right){\rm CNOT}_{ty}\left(I_{t}\otimes{\rm PBS}_{xy}\right)\\
 & \times & \left({\rm CNOT}_{tx}\otimes I_{y}\right)\,\frac{-1}{\sqrt{2}}\left[|1\rangle|0\rangle|{\rm vac}\rangle+|1\rangle|{\rm vac}\rangle|0\rangle\right]_{txy}\\
 & = & \left(I_{t}\otimes H_{x}\otimes H_{y}\right)\left({\rm CNOT}_{tx}\otimes I_{y}\right){\rm CNOT}_{ty}\left(I_{t}\otimes{\rm PBS}_{xy}\right)\,\frac{-1}{\sqrt{2}}\left[|1\rangle|1\rangle|{\rm vac}\rangle+|1\rangle|{\rm vac}\rangle|0\rangle\right]_{txy}\\
 & = & \left(I_{t}\otimes H_{x}\otimes H_{y}\right)\left({\rm CNOT}_{tx}\otimes I_{y}\right){\rm CNOT}_{ty}\,\frac{-1}{\sqrt{2}}\left[|1\rangle|1\rangle|{\rm vac}\rangle+|1\rangle|0\rangle|{\rm vac}\rangle\right]_{txy}\\
 & = & \left(I_{t}\otimes H_{x}\otimes H_{y}\right)\left({\rm CNOT}_{tx}\otimes I_{y}\right)\,\frac{-1}{\sqrt{2}}\left[|1\rangle|1\rangle|{\rm vac}\rangle+|1\rangle|0\rangle|{\rm vac}\rangle\right]_{txy}\\
 & = & \left(I_{t}\otimes H_{x}\otimes H_{y}\right)\,\frac{-1}{\sqrt{2}}\left[|1\rangle|0\rangle|{\rm vac}\rangle+|1\rangle|1\rangle|{\rm vac}\rangle\right]_{txy}\\
 & = & \frac{-1}{2}\left[|1\rangle\left(|0\rangle+|1\rangle\right)|{\rm vac}\rangle+|1\rangle\left(|0\rangle-|1\rangle\right)|{\rm vac}\rangle\right]_{txy}\\
 & = & -|1\rangle_{t}|0\rangle_{x}|{\rm vac}\rangle_{y}
\end{array}.\label{eq:Final_State_Bob_Zero_Alice_One_E3}
\end{equation}
Henceforth, we will only provide the key results. Let's consider the
scenario where Bob prepares $|1\rangle_{t}$, $|+\rangle_{t}$ and
$|-\rangle_{t}$. Following the B-A attack, the composite system becomes,

\[
\begin{array}{lcl}
|{\rm B-A}\rangle_{|1\rangle E_{3}} & = & \frac{1}{\sqrt{2}}\left[|1\rangle|{\rm vac}\rangle|0\rangle+|1\rangle|1\rangle|{\rm vac}\rangle\right]_{txy}\end{array},
\]

\[
\begin{array}{lcl}
|{\rm B-A}\rangle_{|+\rangle E_{3}} & = & \frac{1}{2}\left[|0\rangle|0\rangle|{\rm vac}\rangle+|0\rangle|{\rm vac}\rangle|1\rangle+|1\rangle|{\rm vac}\rangle|0\rangle+|1\rangle|1\rangle|{\rm vac}\rangle\right]_{txy}\end{array},
\]
and

\[
\begin{array}{lcl}
|{\rm B-A}\rangle_{|-\rangle E_{3}} & = & \frac{1}{2}\left[|0\rangle|0\rangle|{\rm vac}\rangle+|0\rangle|{\rm vac}\rangle|1\rangle-|1\rangle|{\rm vac}\rangle|0\rangle-|1\rangle|1\rangle|{\rm vac}\rangle\right]_{txy}\end{array},
\]
respectively.

The composite system after Eve's attack operation $Q_{txy}^{\prime-1}$
(A-B attack) is as follows: When Alice encodes $1$ by applying the
operation $iY_{t}^{1}$ to the Bob's initial states $|1\rangle_{t}$,
$|+\rangle_{t}$ and $|-\rangle_{t}$,

\begin{equation}
\begin{array}{lcl}
|{\rm A-B}\rangle_{|1\rangle E_{3}}^{1} & = & |0\rangle_{t}|0\rangle_{x}|{\rm vac}\rangle_{y}\end{array},\label{eq:Final_State_Bob_One_Alice_One_E3}
\end{equation}

\begin{equation}
\begin{array}{lcl}
|{\rm A-B}\rangle_{|+\rangle E_{3}}^{1} & = & |-\rangle_{t}|0\rangle_{x}|{\rm vac}\rangle_{y}\end{array},\label{eq:Final_State_Bob_Plus_Alice_One_E3}
\end{equation}
and

\begin{equation}
\begin{array}{lcl}
|{\rm A-B}\rangle_{|-\rangle E_{3}}^{1} & = & -|+\rangle_{t}|0\rangle_{x}|{\rm vac}\rangle_{y}\end{array},\label{eq:Final_State_Bob_Minus_Alice_One_E3}
\end{equation}
respectively.

The final composite systems resulting from Eve's two-way attacks (B-A
attack and A-B attack) when Alice encodes $0$ (using the $I_{t}$
operation) with Bob's initial states, $|0\rangle_{t}$, $|1\rangle_{t}$,
$|+\rangle_{t}$ and $|-\rangle_{t}$ are,

\begin{equation}
\begin{array}{lcl}
|{\rm A-B}\rangle_{|0\rangle E_{3}}^{0} & = & |0\rangle_{t}|{\rm vac}\rangle_{x}|0\rangle_{y}\end{array},\label{eq:Final_State_Bob_Zero_Alice_Zero_E3}
\end{equation}

\begin{equation}
\begin{array}{lcl}
|{\rm A-B}\rangle_{|1\rangle E_{3}}^{0} & = & |1\rangle_{t}|{\rm vac}\rangle_{x}|0\rangle_{y}\end{array},\label{eq:Final_State_Bob_One_Alice_Zero_E3}
\end{equation}

\begin{equation}
\begin{array}{lcl}
|{\rm A-B}\rangle_{|+\rangle E_{3}}^{0} & = & |+\rangle_{t}|{\rm vac}\rangle_{x}|0\rangle_{y}\end{array},\label{eq:Final_State_Bob_Plus_Alice_Zero_E3}
\end{equation}
and

\begin{equation}
\begin{array}{lcl}
|{\rm A-B}\rangle_{|-\rangle E_{3}}^{0} & = & |-\rangle_{t}|{\rm vac}\rangle_{x}|0\rangle_{y}\end{array},\label{eq:Final_State_Bob_Minus_Alice_Zero_E3}
\end{equation}
respectively.

\emph{Eve's optimal strategy }Eve's decoding process is characterized
by two scenarios: one in which she decodes $k=0$ when her ancillary
spatial modes $x$ and $y$ are in the empty state and $|0\rangle$
state, respectively, i.e., $|{\rm vac}\rangle_{x}|0\rangle_{y}$,
and $k=1$ when she decodes with the $x$ mode in $|0\rangle$ state
and $y$ mode in empty state, i.e., $|0\rangle_{x}|{\rm vac}\rangle_{y}$.
These scenarios are described using a parameter $p_{jmk}^{E_{3}}$,
which quantifies the joint probability, where $j$, $m$ and $k$
represent Alice, Bob and Eve's encoding, decoding and decoding information,
respectively. We can write the joint probabilities using Eqs. (\ref{eq:Final_State_Bob_Zero_Alice_One_E3})
$-$ (\ref{eq:Final_State_Bob_Minus_Alice_Zero_E3}),

\begin{equation}
\begin{array}{lcl}
p_{000}^{E_{3}} & = & q\\
\\
p_{001}^{E_{3}} & = & p_{010}^{E_{3}}=p_{011}^{E_{3}}=0\\
\\
p_{100}^{E_{3}} & = & p_{101}^{E_{3}}=p_{110}^{E_{3}}=0\\
\\
p_{111}^{E_{3}} & = & \left(1-q\right)
\end{array}.\label{eq:Joint_Probability_E3_Attack}
\end{equation}
By utilizing Equation (\ref{eq:Joint_Probability_E3_Attack}), we
can derive the mutual information expression among Alice, Bob and
Eve. This can be represented as follows,

\begin{equation}
\begin{array}{lcl}
H\left({\rm B|A}\right)_{E_{3}} & = & 0\\
\\
H\left({\rm B}\right)_{E_{3}} & = & \mathbf{H}\left[q\right]+\mathbf{H}\left[1-q\right]\\
\\
I\left({\rm A,B}\right)_{E_{3}} & = & \mathbf{H}\left[q\right]+\mathbf{H}\left[1-q\right]
\end{array},\label{eq:Mutual_Information_Alice_Bob_E3_Attack}
\end{equation}

\begin{equation}
\begin{array}{lcl}
H\left({\rm E|A}\right)_{E_{3}} & = & 0\\
\\
H\left({\rm E}\right)_{E_{3}} & = & \mathbf{H}\left[q\right]+\mathbf{H}\left[1-q\right]\\
\\
I\left({\rm A,E}\right)_{E_{3}} & = & \mathbf{H}\left[q\right]+\mathbf{H}\left[1-q\right]
\end{array},\label{eq:Mutual_Information_Alice_Eve_E3_Attack}
\end{equation}
and

\begin{equation}
\begin{array}{lcl}
H\left({\rm E|B}\right)_{E_{3}} & = & 0\\
\\
H\left({\rm E}\right)_{E_{3}} & = & \mathbf{H}\left[q\right]+\mathbf{H}\left[1-q\right]\\
\\
I\left({\rm B,E}\right)_{E_{3}} & = & \mathbf{H}\left[q\right]+\mathbf{H}\left[1-q\right]
\end{array}.\label{eq:Mutual_Information_Bob_Eve_E3_Attack}
\end{equation}
It is also evident that eavesdropping causes QBER,

\begin{equation}
\begin{array}{lcl}
{\rm QBER}_{E_{3}} & = & \underset{j\ne m}{\sum}p_{jmk}^{E_{3}}\\
 & = & 0
\end{array}.\label{eq:QBER_E3_Attack}
\end{equation}

\textbf{\emph{Control mode:}} In the considered scenario, when Bob
prepares the state $|0\rangle_{t}$ and Alice subsequently measures
it in the $Z$ basis, the composite system $|B-A\rangle_{|0\rangle E_{3}}$
collapses into one of two possible states with equal probability:
$|0\rangle_{t}|0\rangle_{x}|{\rm vac}\rangle_{y}$ or $|0\rangle_{t}|{\rm vac}\rangle_{x}|1\rangle_{y}$.
Following Eve's $Q_{txy}^{\prime-1}$ attack (A-B attack), the composite
systems undergo further transformations,

\begin{equation}
\begin{array}{lcl}
Q_{txy}^{\prime-1}\,|0\rangle_{t}|0\rangle_{x}|{\rm vac}\rangle_{y} & = & \frac{1}{\sqrt{2}}\left[|0\rangle|{\rm vac}\rangle|0\rangle+|0\rangle|{\rm vac}\rangle|1\rangle\right]_{txy}\\
\\
Q_{txy}^{\prime-1}\,|0\rangle_{t}|{\rm vac}\rangle_{x}|1\rangle_{y} & = & \frac{1}{\sqrt{2}}\left[|0\rangle|{\rm vac}\rangle|0\rangle-|0\rangle|{\rm vac}\rangle|1\rangle\right]_{txy}
\end{array}\label{eq:Control_E3_Bob_Zero_State}
\end{equation}
When Bob's initial state is $|1\rangle_{t}$, and Alice measures in
the $Z$ basis, the composite system $|B-A\rangle_{|1\rangle E_{3}}$
collapses to the state $|1\rangle_{t}|{\rm vac}\rangle_{x}|0\rangle_{y}$
and $|1\rangle_{t}|1\rangle_{x}|{\rm vac}\rangle_{y}$ with equal
probability. After Eve's $Q_{txy}^{\prime-1}$ attack (A-B attack),
the composite system transforms accordingly,

\begin{equation}
\begin{array}{lcl}
Q_{txy}^{\prime-1}\,|1\rangle_{t}|{\rm vac}\rangle_{x}|0\rangle_{y} & = & \frac{1}{\sqrt{2}}\left[|1\rangle|{\rm vac}\rangle|0\rangle+|1\rangle|{\rm vac}\rangle|1\rangle\right]_{txy}\\
\\
Q_{txy}^{\prime-1}\,|1\rangle_{t}|1\rangle_{x}|{\rm vac}\rangle_{y} & = & \frac{1}{\sqrt{2}}\left[|1\rangle|{\rm vac}\rangle|0\rangle-|1\rangle|{\rm vac}\rangle|1\rangle\right]_{txy}
\end{array}.\label{eq:Control_E3_Bob_One_State}
\end{equation}
Similarly, when Bob's initial state is $|+\rangle_{t}$, and Alice
measures in the $X$ basis, the composite system $|B-A\rangle_{|+\rangle E_{3}}$
collapses to one of the following states with equal probability: $|+\rangle_{t}|0\rangle_{x}|{\rm vac}\rangle_{y}$,
$|-\rangle_{t}|0\rangle_{x}|{\rm vac}\rangle_{y}$, $|+\rangle_{t}|{\rm vac}\rangle_{x}|1\rangle_{y}$,
$|-\rangle_{t}|{\rm vac}\rangle_{x}|1\rangle_{y}$, $|+\rangle_{t}|{\rm vac}\rangle_{x}|0\rangle_{y}$,
$|-\rangle_{t}|{\rm vac}\rangle_{x}|0\rangle_{y}$, $|+\rangle_{t}|1\rangle_{x}|{\rm vac}\rangle_{y}$
and $|-\rangle_{t}|1\rangle_{x}|{\rm vac}\rangle_{y}$. Subsequently,
the composite systems after Eve's $Q_{txy}^{\prime-1}$ attack (A-B
attack) are described,

\begin{equation}
\begin{array}{lcl}
Q_{txy}^{\prime-1}\,|+\rangle_{t}|0\rangle_{x}|{\rm vac}\rangle_{y} & = & \frac{1}{2\sqrt{2}}\left[\left(|+\rangle+|-\rangle\right)_{t}|{\rm vac}\rangle_{x}\left(|0\rangle+|1\rangle\right)_{y}+\left(|+\rangle-|-\rangle\right)_{t}\left(|0\rangle+|1\rangle\right)_{x}|{\rm vac}\rangle_{y}\right]\\
\\
Q_{txy}^{\prime-1}\,|+\rangle_{t}|{\rm vac}\rangle_{x}|1\rangle_{y} & = & \frac{1}{2\sqrt{2}}\left[\left(|+\rangle+|-\rangle\right)_{t}|{\rm vac}\rangle_{x}\left(|0\rangle-|1\rangle\right)_{y}+\left(|+\rangle-|-\rangle\right)_{t}\left(|0\rangle-|1\rangle\right)_{x}|{\rm vac}\rangle_{y}\right]\\
\\
Q_{txy}^{\prime-1}\,|+\rangle_{t}|{\rm vac}\rangle_{x}|0\rangle_{y} & = & \frac{1}{2\sqrt{2}}\left[\left(|+\rangle+|-\rangle\right)_{t}\left(|0\rangle+|1\rangle\right)_{x}|{\rm vac}\rangle_{y}+\left(|+\rangle-|-\rangle\right)_{t}|{\rm vac}\rangle_{x}\left(|0\rangle+|1\rangle\right)_{y}\right]\\
\\
Q_{txy}^{\prime-1}\,|+\rangle_{t}|1\rangle_{x}|{\rm vac}\rangle_{y} & = & \frac{1}{2\sqrt{2}}\left[\left(|+\rangle+|-\rangle\right)_{t}\left(|0\rangle-|1\rangle\right)_{x}|{\rm vac}\rangle_{y}+\left(|+\rangle-|-\rangle\right)_{t}|{\rm vac}\rangle_{x}\left(|0\rangle-|1\rangle\right)_{y}\right]\\
\\
Q_{txy}^{\prime-1}\,|-\rangle_{t}|0\rangle_{x}|{\rm vac}\rangle_{y} & = & \frac{1}{2\sqrt{2}}\left[\left(|+\rangle+|-\rangle\right)_{t}|{\rm vac}\rangle_{x}\left(|0\rangle+|1\rangle\right)_{y}-\left(|+\rangle-|-\rangle\right)_{t}\left(|0\rangle+|1\rangle\right)_{x}|{\rm vac}\rangle_{y}\right]\\
\\
Q_{txy}^{\prime-1}\,|-\rangle_{t}|{\rm vac}\rangle_{x}|1\rangle_{y} & = & \frac{1}{2\sqrt{2}}\left[\left(|+\rangle+|-\rangle\right)_{t}|{\rm vac}\rangle_{x}\left(|0\rangle-|1\rangle\right)_{y}-\left(|+\rangle-|-\rangle\right)_{t}\left(|0\rangle-|1\rangle\right)_{x}|{\rm vac}\rangle_{y}\right]\\
\\
Q_{txy}^{\prime-1}\,|-\rangle_{t}|{\rm vac}\rangle_{x}|0\rangle_{y} & = & \frac{1}{2\sqrt{2}}\left[\left(|+\rangle+|-\rangle\right)_{t}\left(|0\rangle+|1\rangle\right)_{x}|{\rm vac}\rangle_{y}-\left(|+\rangle-|-\rangle\right)_{t}|{\rm vac}\rangle_{x}\left(|0\rangle+|1\rangle\right)_{y}\right]\\
\\
Q_{txy}^{\prime-1}\,|-\rangle_{t}|1\rangle_{x}|{\rm vac}\rangle_{y} & = & \frac{1}{2\sqrt{2}}\left[\left(|+\rangle+|-\rangle\right)_{t}\left(|0\rangle-|1\rangle\right)_{x}|{\rm vac}\rangle_{y}-\left(|+\rangle-|-\rangle\right)_{t}|{\rm vac}\rangle_{x}\left(|0\rangle-|1\rangle\right)_{y}\right]
\end{array}.\label{eq:Control_E3_Bob_Plus_State}
\end{equation}
If Bob's initial state is $|-\rangle_{t}$, the composite state after
Eve's B-A attack is as follows,

\[
\begin{array}{lcl}
|{\rm B-A}\rangle_{|-\rangle E_{3}} & = & \frac{1}{2}\left[|0\rangle|0\rangle|{\rm vac}\rangle+|0\rangle|{\rm vac}\rangle|1\rangle-|1\rangle|{\rm vac}\rangle|0\rangle-|1\rangle|1\rangle|{\rm vac}\rangle\right]_{txy}\\
 & = & \frac{1}{2\sqrt{2}}\left[\left(|+\rangle+|-\rangle\right)\left(|0\rangle|{\rm vac}\rangle+|{\rm vac}\rangle|1\rangle\right)-\left(|+\rangle-|-\rangle\right)\left(|{\rm vac}\rangle|0\rangle-|1\rangle|{\rm vac}\rangle\right)\right]_{txy}
\end{array}.
\]
From an intuitive perspective, it becomes evident that the entire
scenario appears analogous whether Bob's initial state is $|-\rangle_{t}$
or $|+\rangle_{t}$. Therefore, Eve's A-B attack will have the same
impact on all four different outcomes of Alice's measurement, leading
to identical results as described by Eq. (\ref{eq:Control_E3_Bob_Plus_State}).
We get the probability of undetected Eve in the $E_{3}$ attack scenario
using Eqs. (\ref{eq:Control_E3_Bob_Zero_State}) $-$ (\ref{eq:Control_E3_Bob_Plus_State}),

\[
\begin{array}{lcl}
P_{nd}^{E_{3}} & = & \frac{1}{4}\left(1+1+\frac{1}{4}+\frac{1}{4}\right)=\frac{5}{8}\end{array},
\]
the average detection probability of Eve's presence in $E_{3}$ attack
scenario is,

\begin{equation}
\begin{array}{lcl}
P_{d}^{E_{3}} & = & \frac{1}{2}\left(1-\frac{5}{8}\right)=\frac{3}{16}=0.1875\end{array}.\label{eq:Eve_Detection_Probability_E3_Attack}
\end{equation}
Alice and Bob can tolerate the maximum detection probability of Eve's
presence in control mode for $E_{3}$ attack scenario is $0.1875$.
In other words, the threshold bound for the probability of detecting
Eve is $18.75\%$.

\section*{Appendix D}

First, we define Eve's IR attack ($E_{4}$ attack), afterward, we
compute mutual information among Alice, Bob and Eve. As B-A attack,
Eve carries out a uniform (random) measurement with $Z$ and $X$
bases on Bob's qubit and sends the projected qubit to Alice. either
her encoding operation (in message mode) or her measurement operation
(in control mode) on the received qubit and then sends it back to
Bob. In the A-B attack, Eve applies the same measurement operation
to the qubit that she had previously employed in the B-A attack.

\textbf{\emph{Message mode:}} After performing a tedious calculation,
we can obtain the joint probability, $p_{jmk}$, where $j$, $m$
and $k$ represent Alice, Bob and Eve's encoding, decoding and decoding
information, respectively.

\begin{equation}
\begin{array}{lcl}
p_{000}^{E_{4}} & = & \frac{3}{4}q\\
\\
p_{001}^{E_{4}} & = & p_{011}^{E_{4}}=0\\
\\
p_{010}^{E_{4}} & = & \frac{1}{4}q\\
\\
p_{100}^{E_{4}} & = & p_{110}^{E_{4}}=0\\
\\
p_{101}^{E_{4}} & = & \frac{1}{4}\left(1-q\right)\\
\\
p_{111}^{E_{4}} & = & \frac{3}{4}\left(1-q\right)
\end{array}.\label{eq:Joint_Probability_E4_Attack}
\end{equation}
By utilizing Equation (\ref{eq:Joint_Probability_E4_Attack}), we
can derive the mutual information expression among Alice, Bob and
Eve. This can be represented as follows,

\begin{equation}
\begin{array}{lcl}
H\left({\rm B|A}\right)_{E_{4}} & = & 0.811278\\
\\
H\left({\rm B}\right)_{E_{4}} & = & \mathbf{H}\left[\frac{1}{4}\left(1+2q\right)\right]+\mathbf{H}\left[\frac{1}{4}\left(3-2q\right)\right]\\
\\
I\left({\rm A,B}\right)_{E_{4}} & = & H\left({\rm B}\right)_{E_{4}}-H\left({\rm B|A}\right)_{E_{4}}
\end{array},\label{eq:Mutual_Information_Alice_Bob_E4_Attack}
\end{equation}

\begin{equation}
\begin{array}{lcl}
H\left({\rm E|A}\right)_{E_{4}} & = & 0\\
\\
H\left({\rm E}\right)_{E_{4}} & = & \mathbf{H}\left[q\right]+\mathbf{H}\left[\left(1-q\right)\right]\\
\\
I\left({\rm A,E}\right)_{E_{4}} & = & \mathbf{H}\left[q\right]+\mathbf{H}\left[\left(1-q\right)\right]
\end{array},\label{eq:Mutual_Information_Alice_Eve_E4_Attack}
\end{equation}
and

\begin{equation}
\begin{array}{lcl}
H\left({\rm B|E}\right)_{E_{4}} & = & 0.811278\\
\\
H\left({\rm B}\right)_{E_{4}} & = & \mathbf{H}\left[\frac{1}{4}\left(1+2q\right)\right]+\mathbf{H}\left[\frac{1}{4}\left(3-2q\right)\right]\\
\\
I\left({\rm B,E}\right)_{E_{4}} & = & H\left({\rm B}\right)_{E_{4}}-H\left({\rm B|E}\right)_{E_{4}}
\end{array}.\label{eq:Mutual_Information_Bob_Eve_E4_Attack}
\end{equation}
It is also evident that eavesdropping causes QBER,

\begin{equation}
\begin{array}{lcl}
{\rm QBER}_{E_{4}} & = & \underset{j\ne m}{\sum}p_{jmk}^{E_{4}}\\
 & = & \frac{1}{4}q+\frac{1}{4}\left(1-q\right)\\
 & = & 0.25
\end{array}.\label{eq:QBER_E4_Attack}
\end{equation}

\textbf{\emph{Control mode:}} In a two-way BB84-type security setup,
the probability of Eve going undetected during both-way transmission
is as follows,
\[
\begin{array}{lcl}
P_{nd}^{E_{4}} & = & \frac{1}{2}\times\frac{1}{2}=\frac{1}{4}\end{array},
\]
the average detection probability of Eve's presence in the $E_{4}$
attack scenario is,

\begin{equation}
\begin{array}{lcl}
P_{d}^{E_{4}} & = & \frac{1}{2}\left(1-\frac{1}{4}\right)=\frac{3}{8}=0.375\end{array}.\label{eq:Eve_Detection_Probability_E4_Attack}
\end{equation}
Alice and Bob can tolerate the maximum detection probability of Eve's
presence in control mode for $E_{4}$ attack scenario is $0.375$.
In other words, the threshold for the probability of detecting Eve
is $37.5\%$.

\section*{Appendix E}

\begin{table}[h]
\begin{centering}
\begin{tabular}{|c|c|c|c|c|c|}
\hline 
\multirow{4}{*}{$E_{1}$-$E_{2}$ game scenario} & Nash equilibrium point ($p,q,r$)  & Alice's/Bob's payoff  & Eve's payoff  & Payoff difference & $\epsilon_{E_{1}-E_{2}}$\tabularnewline
\cline{2-6} \cline{3-6} \cline{4-6} \cline{5-6} \cline{6-6} 
 & $\left(0.72,0.208,0.225\right)$ & 0.055457 & 0.194543 & 0.13908 & 0.692404\tabularnewline
 & $\left(0.45,0.195,0.005\right)$ & 0.0446318 & 0.205368 & 0.16073 & 0.610303\tabularnewline
 &  &  &  &  & \tabularnewline
\hline 
\multirow{12}{*}{$E_{1}$-$E_{3}$ game scenario} & Nash equilibrium point ($p,q,r$)  & Alice's/Bob's payoff  & Eve's payoff  & Payoff difference & $\epsilon_{E_{1}-E_{3}}$\tabularnewline
\cline{2-6} \cline{3-6} \cline{4-6} \cline{5-6} \cline{6-6} 
 & $\left(0.22,0.716,0.88\right)$ & -0.110497 & 0.360497 & 0.47099 & 0.152451\tabularnewline
 & $\left(0.442,0.75,0.999\right)$ & -0.0862188 & 0.336219 & 0.42243 & 0.18007\tabularnewline
 & $\left(0.41,0.39,0.412\right)$ & -0.157149 & 0.407149 & 0.56429 & 0.177181\tabularnewline
 & $\left(0.76,0.577,0.585\right)$ & -0.136264 & 0.386264 & 0.52252 & 0.21776\tabularnewline
 & $\left(0.56,0.14,0.292\right)$ & -0.0796824 & 0.329682 & 0.40936 & 0.195874\tabularnewline
 & $\left(0.325,0.064,0.532\right)$ & -0.0134987 & 0.263499 & 0.27699 & 0.329893\tabularnewline
 & $\left(0.84,0.047,0.525\right)$ & 0.0324084 & 0.217592 & 0.18518 & 0.460299\tabularnewline
 & $\left(0.485,0.465,0.915\right)$ & -0.090828 & 0.340828 & 0.43165 & 0.363472\tabularnewline
 & $\left(0.235,0.096,0.83\right)$ & -0.013356 & 0.263356 & 0.27671 & 0.463323\tabularnewline
 & $\left(0.47,0.195,0.93\right)$ & -0.0182231 & 0.268223 & 0.28644 & 0.550258\tabularnewline
 &  &  &  &  & \tabularnewline
\hline 
\multirow{11}{*}{$E_{2}$-$E_{3}$ game scenario} & Nash equilibrium point ($p,q,r$)  & Alice's/Bob's payoff  & Eve's payoff  & Payoff difference  & $\epsilon_{E_{2}-E_{3}}$\tabularnewline
\cline{2-6} \cline{3-6} \cline{4-6} \cline{5-6} \cline{6-6} 
 & $\left(0.385,0.215,0.262\right)$ & -0.111965 & 0.361965 & 0.47393 & 0.151087\tabularnewline
 & $\left(0.47,0.055,0.205\right)$ & -0.0276507 & 0.277651 & 0.3053 & 0.143882\tabularnewline
 & $\left(0.25,0.096,0.54\right)$ & -0.0216673 & 0.271667 & 0.29633 & 0.31482\tabularnewline
 & $\left(0.24,0.268,0.71\right)$ & -0.0436386 & 0.293639 & 0.33727 & 0.35838\tabularnewline
 & $\left(0.70,0.138,0.58\right)$ & -0.00442078 & 0.254421 & 0.25884 & 0.430969\tabularnewline
 & $\left(0.284,0.02,0.472\right)$ & 0.0188573 & 0.231143 & 0.21228 & 0.298653\tabularnewline
 & $\left(0.235,0.02,0.758\right)$ & 0.0320242 & 0.217976 & 0.18595 & 0.461603\tabularnewline
 & $\left(0.222,0.10,0.865\right)$ & 0.015688 & 0.234312 & 0.21862 & 0.492488\tabularnewline
 & $\left(0.54,0.048,0.795\right)$ & 0.0558727 & 0.194127 & 0.13825 & 0.587155\tabularnewline
 & $\left(0.80,0.115,0.885\right)$ & 0.0722149 & 0.177785 & 0.10557 & 0.709991\tabularnewline
\hline 
\multirow{10}{*}{$E_{1}$-$E_{4}$ game scenario} & Nash equilibrium point ($p,q,r$)  & Alice's/Bob's payoff  & Eve's payoff  & Payoff difference  & $\epsilon_{E_{1}-E_{4}}$\tabularnewline
\cline{2-6} \cline{3-6} \cline{4-6} \cline{5-6} \cline{6-6} 
 & $(0.23,0.095,0.825)$ & -0.00433851 & 0.254339 & 0.25867 & 0.502924\tabularnewline
 & $(0.245,0.008,0.76)$ & 0.0492999 & 0.2007 & 0.1514 & 0.529315\tabularnewline
 & $(0.572,0.02,0.765)$ & 0.0750153 & 0.174985 & 0.0999 & 0.648014\tabularnewline
 & $(0.928,0.032,0.774)$ & 0.0997124 & 0.150288 & 0.0505 & 0.77876\tabularnewline
 & $(0.324,0.065,0.535)$ & 0.0114311 & 0.238569 & 0.22713 & 0.447399\tabularnewline
 & $(0.85,0.045,0.522)$ & 0.0603314 & 0.189669 & 0.12933 & 0.580622\tabularnewline
 & $(0.405,0.387,0.415)$ & -0.124349 & 0.374349 & 0.49869 & 0.324962\tabularnewline
 & $(0.54,0.15,0.295)$ & -0.0471361 & 0.297136 & 0.34427 & 0.369328\tabularnewline
 & $(0.75,0.57,0.582)$ & -0.114078 & 0.364078 & 0.47815 & 0.323478\tabularnewline
\hline 
\end{tabular}
\par\end{centering}
\caption{\label{tab:NE_Points_Payoff}This table provides information on the
Nash equilibrium points for various game scenarios and the qualitative
payoff values for all parties.}

\end{table}

\end{document}